\documentclass[aps,prd,twocolumn,superscriptaddress,nofootinbib]{revtex4-1}

\usepackage{float}
\usepackage{amsfonts}
\usepackage{psfrag}
\usepackage{graphicx}
\usepackage{grffile}
\usepackage{cancel}
\usepackage{natbib}
\usepackage{xr}
\usepackage{hyperref}
\usepackage{verbatim}
\usepackage{epstopdf}
\usepackage{multirow}
\usepackage{commath}
\usepackage{empheq}
\usepackage{placeins}
\usepackage{ragged2e}
\usepackage{amsmath}
\usepackage{gensymb}

\usepackage{booktabs, dcolumn}
\hypersetup{
    unicode=false,          % non-Latin characters in Acrobat’s bookmarks
    pdftoolbar=true,        % show Acrobat’s toolbar?
    pdfmenubar=true,        % show Acrobat’s menu?
    pdffitwindow=false,     % window fit to page when opened
    pdfstartview={FitH},    % fits the width of the page to the window
    pdfauthor={Mariska Hoogkamer},     % author
    colorlinks=true,       % false: boxed links; true: colored links
    linkcolor=blue,          % color of internal links
    citecolor=blue        % color of links to bibliography
}
\newcolumntype{d}[1]{D{.}{.}{#1}}

\RequirePackage[normalem]{ulem} 
\RequirePackage{color}\definecolor{RED}{rgb}{1,0,0}\definecolor{BLUE}{rgb}{0,0,1}

\providecommand{\DIFaddbegin}{} 
\providecommand{\DIFaddend}{} 
\providecommand{\DIFdelbegin}{} 
\providecommand{\DIFdelend}{}

\providecommand{\DIFaddbeginFL}{} 
\providecommand{\DIFaddendFL}{} 
\providecommand{\DIFdelbeginFL}{} 
\providecommand{\DIFdelendFL}{}

\newcommand{\DIFscaledelfig}{0.5}
\RequirePackage{settobox} 
\RequirePackage{letltxmacro} 
\newsavebox{\DIFdelgraphicsbox} 
\newlength{\DIFdelgraphicswidth} 
\newlength{\DIFdelgraphicsheight} 
\LetLtxMacro{\DIFOincludegraphics}{\includegraphics} 
\newcommand{\DIFaddincludegraphics}[2][]{{\color{blue}\fbox{\DIFOincludegraphics[#1]{#2}}}} 
\newcommand{\DIFdelincludegraphics}[2][]{
\sbox{\DIFdelgraphicsbox}{\DIFOincludegraphics[#1]{#2}}
\settoboxwidth{\DIFdelgraphicswidth}{\DIFdelgraphicsbox} 
\settoboxtotalheight{\DIFdelgraphicsheight}{\DIFdelgraphicsbox} 
\scalebox{\DIFscaledelfig}{
\parbox[b]{\DIFdelgraphicswidth}{\usebox{\DIFdelgraphicsbox}\\[-\baselineskip] \rule{\DIFdelgraphicswidth}{0em}}\llap{\resizebox{\DIFdelgraphicswidth}{\DIFdelgraphicsheight}{
\setlength{\unitlength}{\DIFdelgraphicswidth}
\begin{picture}(1,1)
\thicklines\linethickness{2pt} 
{\color[rgb]{1,0,0}\put(0,0){\framebox(1,1){}}}
{\color[rgb]{1,0,0}\put(0,0){\line( 1,1){1}}}
{\color[rgb]{1,0,0}\put(0,1){\line(1,-1){1}}}
\end{picture}
}\hspace*{3pt}}} 
} 
\LetLtxMacro{\DIFOaddbegin}{\DIFaddbegin} 
\LetLtxMacro{\DIFOaddend}{\DIFaddend} 
\LetLtxMacro{\DIFOdelbegin}{\DIFdelbegin} 
\LetLtxMacro{\DIFOdelend}{\DIFdelend} 
\DeclareRobustCommand{\DIFaddbegin}{\DIFOaddbegin \let\includegraphics\DIFaddincludegraphics} 
\DeclareRobustCommand{\DIFaddend}{\DIFOaddend \let\includegraphics\DIFOincludegraphics} 
\DeclareRobustCommand{\DIFdelbegin}{\DIFOdelbegin \let\includegraphics\DIFdelincludegraphics} 
\DeclareRobustCommand{\DIFdelend}{\DIFOaddend \let\includegraphics\DIFOincludegraphics} 
\LetLtxMacro{\DIFOaddbeginFL}{\DIFaddbeginFL} 
\LetLtxMacro{\DIFOaddendFL}{\DIFaddendFL} 
\LetLtxMacro{\DIFOdelbeginFL}{\DIFdelbeginFL} 
\LetLtxMacro{\DIFOdelendFL}{\DIFdelendFL} 
\DeclareRobustCommand{\DIFaddbeginFL}{\DIFOaddbeginFL \let\includegraphics\DIFaddincludegraphics} 
\DeclareRobustCommand{\DIFaddendFL}{\DIFOaddendFL \let\includegraphics\DIFOincludegraphics} 
\DeclareRobustCommand{\DIFdelbeginFL}{\DIFOdelbeginFL \let\includegraphics\DIFdelincludegraphics} 
\DeclareRobustCommand{\DIFdelendFL}{\DIFOaddendFL \let\includegraphics\DIFOincludegraphics} 

\newcommand{\Msun}{$M_\odot$}
\begin{document}

\title{Equation-of-state-informed pulse profile modeling}

\author{Mariska Hoogkamer}
\thanks{These authors contributed equally to this work and are the joint lead authors.}
\affiliation{Anton Pannekoek Institute for Astronomy, University of Amsterdam, Science Park 904, 1098XH Amsterdam, the Netherlands}

\author{Nathan Rutherford}
\thanks{These authors contributed equally to this work and are the joint lead authors.}
\affiliation{Department of Physics and Astronomy, University of New Hampshire, Durham, New Hampshire 03824, USA}

\author{Daniela Huppenkothen}
\affiliation{Anton Pannekoek Institute for Astronomy, University of Amsterdam, Science Park 904, 1098XH Amsterdam, the Netherlands}

\author{Benjamin Ricketts}
\affiliation{Anton Pannekoek Institute for Astronomy, University of Amsterdam, Science Park 904, 1098XH Amsterdam, the Netherlands}
\affiliation{SRON, Niel Bohrweg 4, 2033 CA Leiden, the Netherlands}

\author{Anna L. Watts}
\affiliation{Anton Pannekoek Institute for Astronomy, University of Amsterdam, Science Park 904, 1098XH Amsterdam, the Netherlands}

\author{Melissa Mendes}
\affiliation{Technische Universit\"at Darmstadt, Department of Physics, 64289 Darmstadt, Germany}
\affiliation{ExtreMe Matter Institute EMMI, GSI Helmholtzzentrum f\"ur Schwerionenforschung GmbH, 64291 Darmstadt, Germany}
\affiliation{Max-Planck-Institut f\"ur Kernphysik, Saupfercheckweg 1, 69117 Heidelberg, Germany}

\author{Isak Svensson}
\affiliation{Technische Universit\"at Darmstadt, Department of Physics, 64289 Darmstadt, Germany}
\affiliation{ExtreMe Matter Institute EMMI, GSI Helmholtzzentrum f\"ur Schwerionenforschung GmbH, 64291 Darmstadt, Germany}
\affiliation{Max-Planck-Institut f\"ur Kernphysik, Saupfercheckweg 1, 69117 Heidelberg, Germany}

\author{Achim~Schwenk}
\affiliation{Technische Universit\"at Darmstadt, Department of Physics, 64289 Darmstadt, Germany}
\affiliation{ExtreMe Matter Institute EMMI, GSI Helmholtzzentrum f\"ur Schwerionenforschung GmbH, 64291 Darmstadt, Germany}
\affiliation{Max-Planck-Institut f\"ur Kernphysik, Saupfercheckweg 1, 69117 Heidelberg, Germany}

\author{Michael Kramer}
\affiliation{Max-Planck-Institut für Radioastronomie, Auf dem Hügel 69, D-53121 Bonn, Germany}

\author{Kai Hebeler}
\affiliation{Technische Universit\"at Darmstadt, Department of Physics, 64289 Darmstadt, Germany}
\affiliation{ExtreMe Matter Institute EMMI, GSI Helmholtzzentrum f\"ur Schwerionenforschung GmbH, 64291 Darmstadt, Germany}
\affiliation{Max-Planck-Institut f\"ur Kernphysik, Saupfercheckweg 1, 69117 Heidelberg, Germany}

\author{Tuomo Salmi}
\affiliation{Department of Physics, University of Helsinki, P.O. Box 64, FI-00014 University of Helsinki, Finland}

\author{Devarshi Choudhury}
\affiliation{Anton Pannekoek Institute for Astronomy, University of Amsterdam, Science Park 904, 1098XH Amsterdam, the Netherlands}

\begin{abstract}
NICER has enabled mass–radius inferences for pulsars using pulse profile modeling (PPM), providing constraints on the equation of state (EOS) of cold, dense matter. To date, PPM and EOS inference have been carried out as two separate steps, with the former using EOS-agnostic priors. This approach has several drawbacks. Ideally, one would perform a fully hierarchical Bayesian inference where the pulse profile and EOS model parameters are jointly fit, but implementing such a framework is complex and computationally demanding. Here, we present an intermediate solution introducing an EOS-informed prior on mass-radius into the existing PPM pipeline using normalizing flows. By focusing on the parameter space consistent with certain EOSs, this approach both tightens constraints on neutron star parameters while reducing computational costs and requiring minimal additional implementation effort. We test this approach on two pulsars, PSR J0740+6620 and PSR J0437-4715, and with two EOS model families: a model based on the speed of sound inside the neutron star interior (CS) and a piecewise-polytropic (PP) model. Both EOS models implement constraints from chiral effective field theory calculations of dense matter. For both pulsar datasets, the inferred radius credible intervals are narrower than in the EOS-agnostic case, with CS favoring smaller radii and PP favoring larger radii. For PSR J0437-4715, the EOS-informed priors reveal a new, more extreme geometric mode that is statistically favored but physically questionable. Including the PPM posteriors in the subsequent EOS inference further tightens the mass-radius posteriors through the chiral effective field theory constraints. However, there is also a sensitivity to the high-density extensions, where the PP (CS) model produces a shift toward larger (smaller) radii and corresponding stiffening (softening) of the pressure–energy density relation. 
\end{abstract}

\maketitle

\section{Introduction}\label{sec:introduction}

Neutron stars, with their extreme compactness, provide a unique opportunity to study the behavior of dense nuclear matter and constrain the equation of state (EOS) \cite[see, e.g.,][]{Watts2016, Lattimer2016-EOS, Baym2018-EOS, Tolos2020}. Observations of x-ray emission from their surfaces serve as a powerful tool for probing their internal structure, thereby placing constraints on the EOS of cold, dense matter.

The Neutron Star Interior Composition Explorer (NICER; \cite{Gendreau2016-NICER}), onboard the International Space Station, has played a pivotal role in this effort by detecting soft thermal x-rays from a specific class of neutron stars: rotation-powered millisecond pulsars. These pulsars exhibit rotationally modulated, pulsed x-ray emission, believed to originate from hot spots at the magnetic poles where return currents deposit heat \cite[see, e.g.,][]{Ruderman1975-pulsars, Arons1981, Harding2001}. The x-ray signal is analyzed using the pulse profile modeling (PPM) technique \cite[e.g.,][and references therein]{Watts2019-ppm, Bogdanov2019-part2, Bogdanov2021-part3}, which incorporates relativistic effects arising from the neutron star’s rapid spin and intense gravitational field. This approach enables precise measurements of neutron star masses and radii.

To date, NICER has enabled mass and radius inferences for five pulsars: PSR J0030+0451, \cite{Riley2019-J0030-xpsi, Miller2019-J0030, Salmi2023-atmosphere, Vinciguerra2024-J0030}, PSR J0437-4715 \cite{Choudhury2024-J0437}, PSR J0740+6620 \cite{Miller2021-J0740, Riley2021-J0740, Salmi2022-J0740, Salmi2023-atmosphere, Salmi2024-J0740, Dittmann2024-J0740, Hoogkamer2025-J0740}, PSR J1231-1411 \cite{Salmi2024-J1231, Qi2025-J1231}, and PSR J0614-3329 \cite{Mauviard2025}. The resulting mass-radius ($M$–$R$) constraints have been widely used to constrain the EOS of cold, dense matter \cite[see, e.g.,][]{Rutherford24, Koehn24, Huang25, Golomb25, LiJJ24}.

The X-ray Pulse Simulation and Inference \citep[X-PSI;][]{xpsi} code is an open-source software package for PPM and Bayesian statistical inference.\footnote{\url{https://github.com/xpsi-group/xpsi}} In $M–R$ inference to date, X-PSI has adopted an EOS-agnostic approach, following the methodology in, e.g., \cite{Riley2018-EOS, Raaijmakers2018-EOS}, selecting joint flat priors on mass and radius, and modifying the mass prior when informative radio-timing measurements are available. EOS inference, using the open-source software NEoST~\cite{NEoST2025},\footnote{\url{https://github.com/xpsi-group/neost}} is then performed as a separate analysis step. The main motivation for this two-step approach is to provide the community with posterior samples in the $M$–$R$ plane that can be studied using any EOS model afterward. However, this strategy has some practical drawbacks:

\begin{enumerate}
    \item High computational costs: The accuracy and precision of the PPM-inferred neutron star properties depend on factors such as data quality and the availability of prior information. When these are limited, the resulting parameter space becomes broad or difficult to explore efficiently. For instance, inference for PSR J0437-4715 required $\sim783$k core hours to achieve convergence \cite{Choudhury2024-J0437}. 
    \item Sampling of implausible parameter space: The use of flat priors permits sampling of regions, such as very compact solutions, that are already ruled out by physical EOS constraints - yet these regions incur very expensive multiple-image ray-tracing computations. 
    \item Difficulty with multimodality: A source like PSR J0030+0451 \cite{Salmi2023-atmosphere, Vinciguerra2024-J0030} exhibits complex, multimodal solutions that are difficult to sample thoroughly within reasonable compute time with flat priors. Imposing realistic priors that exclude regions already excluded by EOS constraints will focus sampling only on physically meaningful regions of parameter space, and potentially break multimodality and reduce sampling time.
    \item Because $M$ and $R$ are generally correlated with geometry parameters (e.g., spot size and location), imposing more physically meaningful priors will also lead to inferred geometries that are more realistic and better constrained, in contrast to the current implementation.   
\end{enumerate}

In an ideal scenario, one would perform fully hierarchical Bayesian inference where the pulse profile and EOS parameters are jointly fit. While this would solve the aforementioned issues, the implementation of such a framework is very complex, time consuming, and running it could be computationally intractable. 

At present, the X-PSI and NEoST packages are two distinct frameworks, each focusing on different aspects of the full Bayesian model, namely PPM and EOS inference, respectively. PPM is carried out for individual stars inferring parameters such as mass and radius. EOS parameters are introduced only in the second stage, along with stellar central densities, when information from multiple stars is combined. Properly unifying the PPM and EOS inference would substantially increase the number of parameters, as it would require simultaneous inference of neutron star properties, EOS parameters, and central density even for a single star. This framework would eventually make it possible to carry out PPM for multiple stars at the same time. On the other hand, such a joint inference could benefit from Bayesian shrinkage: The requirement that all stars share the same EOS may help break certain parameter degeneracies, potentially making the overall sampling process more efficient.

As an intermediate step, in this work, we introduce an EOS-informed $M$–$R$ prior into the existing PPM pipeline, while leaving the subsequent EOS inference unchanged. This approach requires minimal additional implementation effort but still constrains the PPM step to regions of parameter space consistent with prior EOS information, thereby providing a way to obtain tighter constraints on neutron star parameters through more physically informed priors. In addition, it helps reduce computational costs by limiting the parameter space that must be sampled, which in turn might lead to quicker convergence. Finally, leaving the PPM and EOS inference as two separate steps has the additional advantage of avoiding having to deal with a substantial increase in the number of parameters as would be the case when unifying the two.     

To inform the PPM pipeline of the underlying cold, dense matter EOS prior, we use the two EOS model families considered in \citet{Rutherford24}. At nuclear densities beyond the Baym-Pethick-Sutherland (BPS) crust \citep{Baym71}, these are based on next-to-next-to-next-to-leading-order (N$^3$LO) chiral effective theory calculations of matter in beta equilibrium from \citet{Keller2023} up to $1.5n_0$, where $n_0 = 0.16 \, \mathrm{fm}^{-3}$ is the nuclear saturation density. The EOS prior thus builds in knowledge of the crust and of the EOS at nuclear densities based on our modern understanding of nuclear forces. At higher densities, we use two possible model extensions, based on the speed of sound inside the neutron star interior (CS) \cite{Greif19} or based on a piecewise-polytropic (PP) model \cite{Hebeler2013}. In this work, we explore the impact of both prior EOS ensembles on the PPM-derived $M$–$R$ and geometry posteriors of PSR J0740+6620 and PSR J0437-4715, and the effects of the derived $M$–$R$ posteriors on the subsequent EOS inference analysis. 

The remainder of this paper is organized as follows. In Sec.~\ref{subsec:full hierarchical model}, we present the ideal case of a full hierarchical Bayesian model for the simultaneous inference of neutron star and EOS properties. Section~\ref{subsec:PPM EOS prior and EOS inference} introduces our intermediate approach that incorporates EOS prior information to constrain neutron star properties, followed by the EOS inference framework using EOS-prior-informed PPM $M–R$ posteriors. Sections~\ref{subsec:xpsi} and \ref{subsec:normflows} describe the X-PSI PPM analysis pipeline and the use of normalizing flows to model the $M–R$ prior derived from the two EOS families. In Sec.~\ref{sec:data}, we summarize the datasets for PSR J0740+6620 and PSR J0437–4715, and outline the prior intervals for the PP and CS EOS model parameters used in this analysis. Section~\ref{sec:Results} presents the EOS-prior-informed PPM posteriors and the resulting constraints on the cold, dense matter EOS. Finally, in Sec.~\ref{sec:Discussion}, we discuss our conclusions and directions for future work.

\section{Methodology}\label{sec:Methodology}

Section~\ref{subsec:full hierarchical model} outlines the ideal scenario of a full hierarchical Bayesian model for the joint inference of neutron star and EOS parameters. In Sec.~\ref{subsec:PPM EOS prior and EOS inference}, we introduce our intermediate method---that does not require substantial implementation effort---which incorporates EOS prior information to constrain neutron star properties and enables EOS inference based on EOS-prior-informed PPM $M–R$ posteriors. Sections~\ref{subsec:xpsi} and \ref{subsec:normflows} detail the X-PSI PPM analysis pipeline and the implementation of normalizing flows to represent the $M–R$ prior derived from the two EOS models, respectively. Complete information of each PPM and EOS inference run, including the exact X-PSI and NEoST versions, data products, posterior sample files, and all the analysis files are available in a Zenodo repository \cite{HoogkamerRutherford2025-Zenodo}.\footnote{\label{zenodo link}\url{https://doi.org/10.5281/zenodo.17257402}}

\subsection{Full hierarchical bayesian model}\label{subsec:full hierarchical model}

\begin{figure}[t!]
\centering
\includegraphics[width=0.9\columnwidth]
{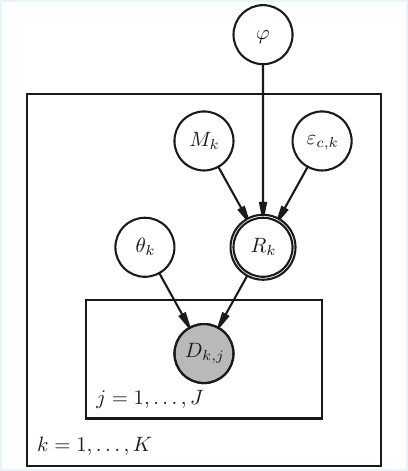}  
\caption{Probabilistic graphical model illustrating the hierarchical Bayesian framework for the joint inference of neutron star properties and EOS parameters. Arrows denote statistical dependences between variables. The shaded gray circle represents observed variables, while double circles indicate deterministic variables. The box represents a summation that applies to all parameters contained within it. See Table~\ref{tab:params hierarical model} for a detailed description of all variables.}
\label{fig:pgm}
\end{figure}

Currently, posterior densities are derived for the parameters of each neutron star independently, and later, the $M$-$R$ posteriors are used as ``data'' in EOS inference. However, under the assumption that all neutron stars adhere to the same EOS, it is much more informative to infer both global EOS parameters and the local parameters for all available neutron stars simultaneously, using a hierarchical Bayesian model. A visual representation of this full model, using probabilistic graphical models, is shown in Fig.~\ref{fig:pgm}. A key advantage of this approach is the occurrence of Bayesian shrinkage: By constraining all neutron stars to lie on a common EOS curve, we should obtain tighter constraints on the $M–R$ relation and the EOS parameters.

We begin by representing the pulse profile data as $\left\{ D_{k,j}\right\}^{K,J}_{k,j=1}$, where $D_{k,j}$ denotes the observed data for the $k$th neutron star with $K$ the total number of neutron stars included in the analysis, and the $j$th data point with $J$ the total number of data points in each pulse profile. See Table~\ref{tab:params hierarical model} for an overview of all parameter definitions.

Next, we specify the prior distribution for the parameters not directly affected by the EOS, such as the inclination angle and the hot spot geometry parameters:
\begin{equation}\label{eq:prior theta}
    p(\left\{ \theta_k\right\}^K_{k=1})=\prod_{k=1}^{K}p(\theta_k) \,,
\end{equation}
with the parameter $\theta_k$ not affected by the EOS, assuming independence between different neutron star sources. 

\begin{table}[t!]
\begin{tabular}{ll}
\hline
\textbf{Parameter} & \textbf{Description} \\ \hline
$\left\{ D_{k,j}\right\}^{K,J}_{k,j=1}$ & \begin{tabular}[c]{@{}l@{}}Observed pulse profile data for neutron stars \\ $k=1, ... , K$ for each data point $j=1, ... , J$\end{tabular} \\
$\left\{ \theta_k\right\}^K_{k=1}$ & Parameters not affected by the EOS \\
$\left\{M_k \right\}^K_{k=1}$ & Gravitational mass for each neutron star $k$ \\
$\left\{R_k \right\}^K_{k=1}$ & Equatorial radius for each neutron star $k$ \\
$\varphi$ & EOS parameters \\
$\left\{ \varepsilon_{c,k}\right\}^K_{k=1}$ & Central energy density for each neutron star $k$ \\
\hline
\end{tabular}
\caption{Definitions for each parameter of the statistical model.}
\label{tab:params hierarical model}
\end{table}

In contrast, the neutron star parameters that are affected by the EOS are governed by a deterministic relationship. Using ($\varphi, \varepsilon_{c,k}$), the vector containing the EOS parameters and the central density of the $k$th neutron star, respectively,  and solving the Tolman-Oppenheimer-Volkoff \citep[TOV; see][]{Tolman1939,Oppenheimer1939} equations, the radius $R_k$ is determined by the $M$–$R$ relationship of the EOS for a given sample from the prior for $M_k$. $M_k$ is sampled from a mass prior based on radio-timing observations (when available), or from a uniform prior (when such observations are unavailable), and $R_k$ is determined using $\varphi$, $\varepsilon_{c,k}$ and $M_k$. 
In this setup, the mass $M_k$ is treated as an independent variable, and we can express its prior as 
\begin{equation}\label{eq:prior m}
    p(\left\{M_k \right\}^K_{k=1})=\prod_{k=1}^{K}p(M_k) \,,
\end{equation}
assuming independence between different neutron stars. 

The prior on the EOS parameters $\varphi$ is given by
\begin{equation}\label{eq:prior varphi}
    p(\varphi) \,.
\end{equation}
Similarly, the prior on the central energy density $\varepsilon_{c,k}$ is given by
\begin{equation}\label{eq:prior rho_c}
    p(\left\{\varepsilon_{c,k} \right\}^K_{k=1})=\prod_{k=1}^{K}p(\varepsilon_{c,k})\,,
\end{equation}
again assuming independence across neutron stars. We note that, strictly speaking, $\left\{\theta_{k} \right\}^K_{k=1}$, $\left\{\varepsilon_{c,k} \right\}^K_{k=1}$, and $\varphi$ depend on certain fixed hyperparameters.\footnote{\label{hyperparams}Hyperparameters are parameters that determine the shape of the prior distribution for a given model parameter. For example, if the prior for an inclination angle follows a normal distribution, the hyperparameters would be its mean and standard deviation.} However, since these hyperparameters are fixed and therefore not random variables, we have chosen not to write them down for clarity. Similarly, all distributions in the joint PPM-EOS inference should be read as conditional on the choices of PPM model \citep[see, e.g.,][]{Riley2019-J0030-xpsi, Vinciguerra2023-J0030-sim} and on the choice of EOS model \citep[see][for the PP and CS models, respectively]{Hebeler2013, Greif19}. To avoid cluttering the notation, we have omitted including these in the conditional densities as well. 

The likelihood is described by
\begin{equation}\label{eq:dream likelihood}\begin{split}
    p(\left\{ D_{k,j}\right\}^{K,J}_{k,j=1}|\left\{ \theta_k, M_k, \varepsilon_{c,k} \right\}^K_{k=1}, \varphi) \\
    =\prod_{k=1}^{K}\prod_{j=1}^{J}p(D_{k,j}|\theta_k, M_k, \varepsilon_{c,k}, \varphi) \,, 
\end{split}\end{equation}
where we assume that both neutron stars and data points within each pulse profile are statistically independent.
By combining Eqs.~\eqref{eq:prior theta}-\eqref{eq:dream likelihood}, we obtain the posterior probability distribution: 
\begin{align}
    p(\left\{\theta_k, M_k, \varepsilon_{c,k} \right\}^K_{k=1}, \varphi|\left\{ D_{k,j}\right\}^{K,J}_{k,j=1}) \nonumber \\
    \propto \prod_{k=1}^{K} \left[\prod_{j=1}^{J}p(D_{k,j}|\theta_k, M_k, \varepsilon_{c,k}, \varphi)\right] \nonumber \\ 
    \times~p(\theta_k) p(M_k) p(\varepsilon_{c,k}) p(\varphi) \label{eq:dream posterior} \,.
\end{align}
As mentioned before, since the equatorial radius $R_k$ of a given source is completely determined by the combination of $(\varphi, \varepsilon_{c,k}, M_k)$, the posterior distribution on $R_k$ can be computed once the posteriors on $\varphi$, $\varepsilon_{c,k}$, and $M_k$ are determined. 

\subsection{Intermediate step: PPM including EOS prior information and EOS inferencing}\label{subsec:PPM EOS prior and EOS inference}

At present, the full hierarchical Bayesian framework described in Sec.~\ref{subsec:full hierarchical model} is computationally intractable. We therefore leave its construction for future work. In this work, we adopt an intermediate approach: first performing PPM with X-PSI, including EOS $M$–$R$ prior information corresponding to a specific EOS family, and then carrying out the subsequent EOS inference using NEoST. In the PPM stage, we restrict ourselves to a family of EOSs, such that the prior constrains the PPM parameter space to only those combinations permitted within that family. While this builds in physics constraints on the EOS, from chiral effective field theory ($\chi$EFT) calculations at nuclear densities, it also introduces model choices from the high-density extensions. During the EOS inference stage, we then infer which of the parameter combinations within that family are most consistent with the data. This approach has the advantage of being quick to implement while still yielding tighter constraints on neutron star parameters by restricting the parameter space to regions consistent with certain EOSs, while also reducing computational cost.

To construct the EOS-informed posterior distribution for the derived neutron star properties, we begin by formulating the posterior obtained via X-PSI. In previous X-PSI analyses of rotation-powered millisecond pulsars \cite[e.g.,][]{Riley2019-J0030-xpsi, Salmi2023-atmosphere, Vinciguerra2024-J0030, Riley2021-J0740, Salmi2022-J0740, Salmi2023-atmosphere, Salmi2024-J0740, Hoogkamer2025-J0740, Choudhury2024-J0437}, analysis was done for a single star ($K=1$) and the EOS prior was not incorporated. Instead, the posterior was given by: 
\begin{equation}
    p(\theta, R, M|\left\{ D_{j}\right\}^{J}_{j=1}) 
    \propto \prod_{j=1}^{J}p(D_{j}|\theta,R, M) p({\theta)p({R) p(M)}} \,. \label{eq:PPM posterior wo EOS}
\end{equation}
where no explicit EOS model was used to constrain the $M–R$ relation.

\begin{figure}[t!]
    \centering
    \includegraphics[width=\columnwidth]{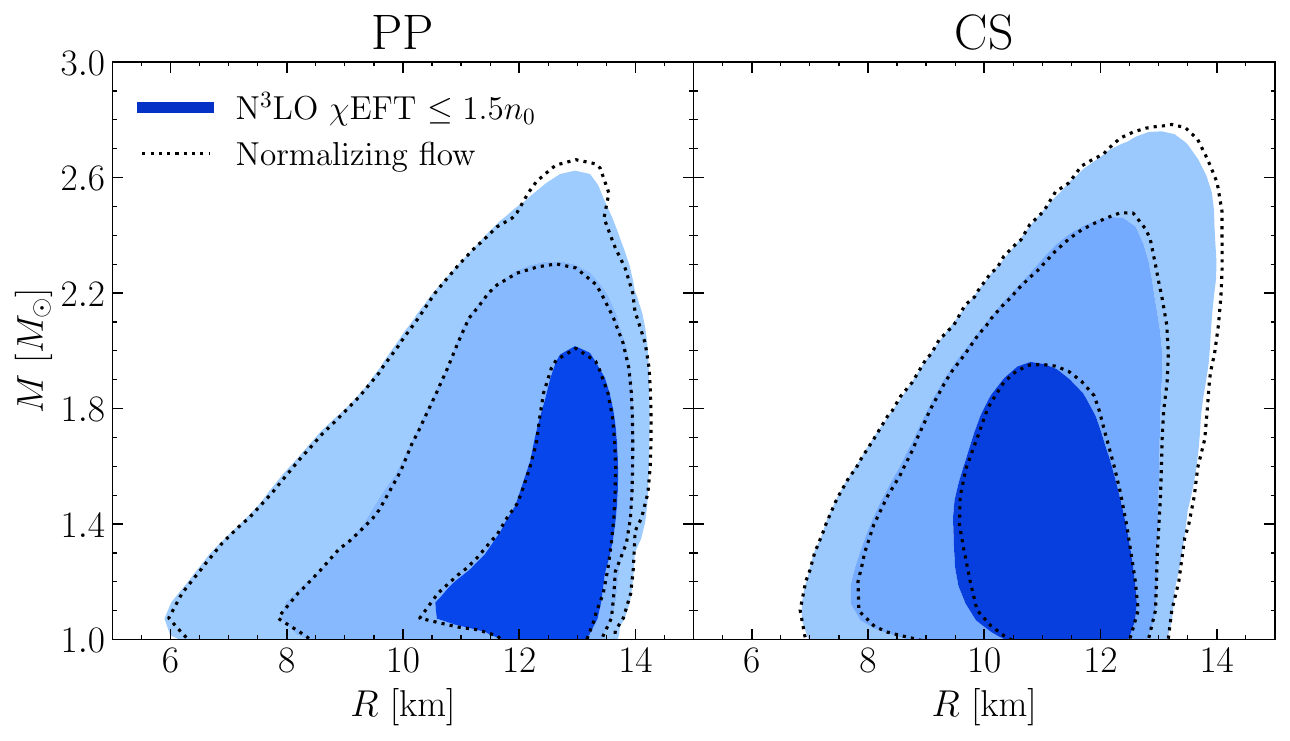}
    \caption{Mass-radius prior distributions for the PP EOS model (left) and CS EOS model (right) when using the N$^3$LO $\chi$EFT calculations up to $1.5 n_0$. The blue contours show the 68\%, 95\%, and 99.7\% credible regions of the EOS prior distributions. Black dotted lines show the corresponding contours from the normalizing flow distributions.}
    \label{fig:MR_priors}
\end{figure}

To fold in the EOS prior information into the $M$–$R$ space within X-PSI, we modify Eq.~\eqref{eq:PPM posterior wo EOS} to define a new posterior:
\begin{align}
    p\big(&\theta, R, M|\left\{ D_{j}\right\}^{J}_{j=1}, \varphi\big) \nonumber \\
    &\propto \prod_{j=1}^{J}p(D_{j}|\theta,\varphi, R, M) p(\theta) p\big(R|\varphi\big) p\big(M\big) p\big(\varphi \big) \,, \label{eq:PPM posterior with EOS}
\end{align}
where the EOS parameters $\varphi$ now inform the $M$–$R$ distribution. In practice, we adopt the following procedure. First, we sample the priors on the EOS parameters and compute their corresponding priors in the $M$–$R$ space using the NEoST framework (described in more detail below), as shown in Fig.~\ref{fig:MR_priors}. We then use these samples to train a normalizing flow (see Sec.~\ref{subsec:normflows}) that approximates the EOS $M$–$R$ prior. Subsequently, if no mass prior is available from radio timing, we draw the mass $M$ from a flat prior. If a mass prior from radio timing is available, we instead sample a mass $M$ from that prior. In either case, we then draw a corresponding radius sample $R$ from the EOS $M$-$R$ prior conditional on the sampled mass $M$.

For the EOS inference, we implement the \citet{Raaijmakers21} Bayesian framework using NEoST (\texttt{v2.1.0})\cite{Raaijmakers2025}.\footnote{The prerelease form of NEoST was used in \citep{Greif19,Raaijmakers19,Raaijmakers20,Raaijmakers21,Rutherford23}. The released version of NEoST has been used in \cite{Rutherford24, Rutherford2024b}.} NEoST uses \texttt{MultiNest} to sample from the chosen EOS parameter space and combines constraints from both the $M–R$ relations derived from PPM and the $M–\Lambda$ relations from gravitational-wave observations. In this work, however, we focus solely on incorporating EOS prior information into the PPM-derived $M–R$ posteriors, and therefore omit the gravitational-wave constraints to highlight the differences between the original and new frameworks. 
Since the EOS model parameters considered in this work---namely PP and CS---are uniformly sampled, the EOS inference framework is identical to that used in our previous work \cite{Raaijmakers21,Rutherford24}. 

The two-step process implemented here naturally raises concerns about double counting the EOS prior, since it is used both to define priors for $M$–$R$ during the PPM stage and again in the subsequent EOS inference. This is especially relevant because uniform priors on the EOS parameters lead to nonuniform priors on neutron star mass and radius due to the nonlinearity of the TOV equations. However, since these equations are deterministic, the transition from EOS parameters to $M$–$R$ space corresponds to a change of variables, which is a common operation in Bayesian inference especially when it reduces the computational costs of sampling. Since in our case the prior on the EOS parameters is flat, applying them during both the PPM and subsequently the EOS inference does not distort the overall shape of the EOS posterior, and thus does not influence EOS parameter estimation. It is important to note, however, that Bayesian model comparison via posterior odds should not be performed in this manner, as the inclusion of the prior in both inference steps would indeed distort the results. 

Therefore, the posterior distribution of the EOS parameters $\varphi$ and central densities $\varepsilon_c$ remains unchanged and is given by
\begin{align}
    p(\bm{\varphi}, \bm{\varepsilon_{c,k}} \,|\, \bm{d})
    \propto 
    p(\bm{\varphi})
    p(\bm{\varepsilon_{c,k}} \,|\, \bm{\varphi})
    \prod_{k=1}^{K} p(M_k, R_k \,|\, \bm{d}_k) \,,
\end{align}
where $\bm{d}$ is the dataset used to constrain the EOS. In this work, $\bm{d}$ consists of the $M–R$ PPM posteriors for PSR J0740+6620 and PSR J0437-4715, obtained using an EOS-informed $M–R$ prior together with a mass prior derived from radio timing. We do not include other data, such as gravitational-wave observations, in the EOS inference.
 
It is important to note that when an EOS-informed prior is adopted in the PPM step, the EOS inference is only self-consistent if the same EOS model is used in both PPM and EOS inference. EOS inference with a different EOS model family from that used to define $M$–$R$ priors for PPM is still possible, provided that the EOS model used to define the PPM priors encompasses the full parameter range of the EOS model applied during the EOS inference. In this scenario, EOS inference is akin to using PPM results derived using EOS-agnostic priors.\footnote{It is important to note that applying different EOS models between the PPM and EOS analysis stages is dependent on whether or not the priors on the EOS model used in the later stage are uniform. If the priors for the EOS model parameters are flat, similar to what is done in this work, then the EOS inference will not be affected by the shape of those priors. However, if the EOS priors are nonuniform, then there is an effective double counting of the prior contribution of the EOS model, thus making the EOS posterior inference statistically inconsistent.} This ensures that the prior informing the PPM-derived $M–R$ distributions---the EOS parameters $\varphi$---remains aligned with the model used in the subsequent EOS inference. 

\subsection{Pulse profile modeling using X-PSI} \label{subsec:xpsi}

We use the open-source X-PSI package, with versions ranging from \verb|v3.0.0| to \verb|v3.0.5| \cite{xpsi}. As done previously \cite[see e.g.,][]{Riley2021-J0740, Salmi2022-J0740, Salmi2024-J0740, Choudhury2024-J0437}, this uses the `Oblate Schwarzschild + Doppler' approximation to model the energy-resolved x-ray pulses from the neutron star \cite{Miller1998, Nath2002, Poutanen2003, Morsink2007, Lo2013, AlGendy2014, Bogdanov2019-part2, Watts2019-ppm}. This approximation takes the oblate shape of the neutron star into account and the relativistic effects resulting from the rapid spin of the rotation-powered millisecond pulsars while treating the exterior spacetime as Schwarzschild. For a more detailed description of X-PSI, see \cite{Riley2019-J0030-xpsi, xpsi}.

X-PSI is used in combination with the nested sampling algorithm \texttt{MultiNest} \cite{Feroz2008-MultimodalNestedSampling, Feroz2009-Multinest, Feroz2019-Multinest} and its Python binding \texttt{PyMultiNest} \cite{Buchner2016-PyMultiNest} to compute posterior samples and calculate the evidence for Bayesian model comparison. We opted for this setup over \texttt{UltraNest} \cite{Buchner2021-UltraNest} due to its lower computational cost, despite \texttt{UltraNest}'s proven convergence and stable results for PSR J0740+6620 \cite{Hoogkamer2025-J0740}. 

For the massive pulsar PSR J0740+6620, we use the exact same setup and sampler settings as in the headline result of \citet{Salmi2024-J0740} and \citet{Hoogkamer2025-J0740}, where NICER and XMM-Newton data are jointly modeled. This setup includes informative priors on mass, inclination, and distance obtained from radio timing \cite{Fonseca2021-J0740}. Section~\ref{subsec:normflows} describes how this mass prior is combined with the EOS-informed $M$–$R$ prior.

For the nearby and bright pulsar PSR J0437–4715, we likewise adopt the same setup as in the headline result of \citet{Choudhury2024-J0437}. This setup incorporates tight informative priors on mass, inclination, and distance based on radio timing \cite{Reardon2024-J0437} (see Sec.~\ref{subsec:normflows} for more details), as well as NICER background models to constrain the nonsource background, cross-validated with XMM-Newton data. Instead of 20k live points and a sampling efficiency of 0.3 used by \citet{Choudhury2024-J0437}, we use 4k live points and 0.1 sampling efficiency for the EOS-informed prior runs to save on computational costs. These settings are expected to be sufficient, as testing the EOS-agnostic case with 4k live points and 0.1 sampling efficiency yielded results consistent with those obtained using 20k live points and 0.3 sampling efficiency in \citet{Choudhury2024-J0437}. 

\subsection{Normalizing flow as EOS prior} \label{subsec:normflows}

Uniform sampling in the EOS space induces a nonuniform prior in the $M$–$R$ space due to the nonlinearity of the TOV equations. Because an analytical transformation of the probability density between these spaces is intractable, we approximate the $M$–$R$ prior to enable sampling from it during the PPM stage. To achieve this, we draw samples from the uniform prior in EOS space, transform them into corresponding pairs in $M$–$R$ space by solving the TOV equations, and then fit a flexible parametric form to the resulting distribution using a normalizing flow. 

Normalizing flows \cite[see, e.g.,][]{Dinh2014, Papamakarios2019, Kobyzev2019} are a class of deep learning models used for density estimation, i.e., learning a function that approximates an unknown probability distribution. Normalizing flows approximate the target distribution by transforming a simple and analytically known distribution, such as a standard normal distribution, into a more complex one by applying a sequence of invertible (i.e., bijective) and differentiable transformations. These transformations are parametrized by neural networks to allow for greater flexibility, and enable both sampling and evaluation of the log-density of the resulting distribution.

In this work, we use normalizing flows to model the EOS $M$–$R$ prior distribution,
\begin{equation} \label{eq:nf}
    p(M,R) = p(R|M)p(M) \,,
\end{equation}
allowing us to efficiently sample joint $M$–$R$ combinations, as well as draw radius samples conditioned on a given mass. The latter is particularly useful in cases where a mass prior is available from radio-timing observations, as for PSR J0740+6620 and PSR J0437-4715. 

In this paper, we use the implementation of normalizing flows provided in \texttt{Pyro} \cite{Pyro2018},\footnote{\url{https://pyro.ai}} with a standard normal distribution $\mathcal{N}(\mu=0, \sigma^2=1)$ as the base distribution. For both mass and radius, we use transformations known as coupling neural spline flows \cite{Durkan2019, Dolatabadi2020, Dinh2014, Muller2018}, specifically element-wise rational-linear spline bijections, where the radius is conditioned on the mass. Each transformation is parametrized by a small dense feedforward neural network that takes $M$–$R$ samples from the EOS prior distributions (see Sec.~\ref{subsec:EOS priors dist}) as input. 
The dataset consists of 100,000 $M$–$R$ samples of which we reserve 70\% for training, 15\% for validation, and another 15\% for testing. Prior to training, the data are standardized using \texttt{Pyro}’s \texttt{StandardScaler} to have zero mean and unit variance.     

To determine the architecture and training hyperparameters,\footnote{In this context, hyperparameters are settings chosen a priori that control how the neural network learns, such as the learning rate and the number of hidden layers.} we perform Bayesian hyperparameter tuning using \texttt{W\&B} \cite{wandb},\footnote{\url{https://wandb.ai}} aiming to select a configuration that balances validation loss (the negative log-likelihood of the samples given the transformation of the distribution) with training efficiency. The sweep varied the hidden layer dimensions ($8~–~40$), the number of transformations  ($1~–~10$) of $p(R|M)$ and $p(M)$ as shown in Eq.~\eqref{eq:nf}, the learning rate ($5\times10^{-4}~-~5\times10^{-2}$), and the number of training epochs ($250~–~12000$ in steps of 250).

\begin{figure}[t!]
    \centering
    \includegraphics[width=\columnwidth]{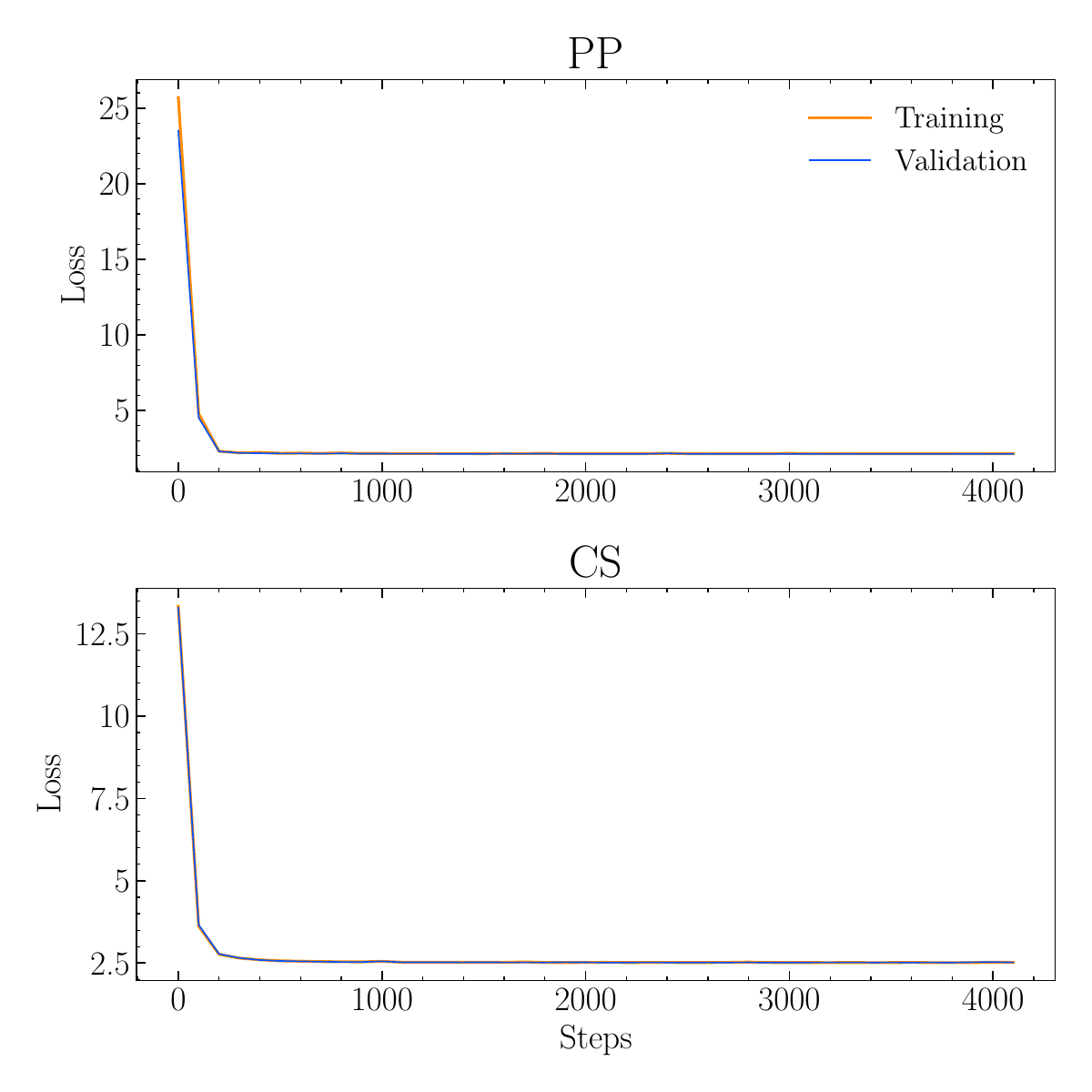}
    \caption{Training (orange) and validation (blue) loss curves for the PP EOS model (top) and CS EOS model (bottom).}
    \label{fig:loss}
\end{figure}

After performing the Bayesian hyperparameter tuning, we selected the best-performing configuration without further exploring alternative setups. Because there were many configurations with similar validation losses, we chose one with a relatively short training time. This configuration---using a normalizing flow with ten transformations for both $p(R|M)$ and $p(M)$---performed well, achieving a test loss of 2.52 for CS and 2.14 for PP (see Fig.~\ref{fig:loss} for the training and validation loss curves) in approximately 27 minutes for both cases. Other configurations with comparable loss values required substantially longer training times. The resulting network contains two hidden layers with 32 neurons each, and ReLU\footnote{ReLU (rectified linear unit) is an activation function defined as $f(x)=\text{max}(0,x)$. Activation functions introduce nonlinearity into the network, allowing the model to capture complex patterns and relationships in the data.} \cite{NairHinton2010} activations following each layer. The normalizing flow is trained on 70,000 $M–R$ samples for 4000 epochs, using the Adam optimizer \cite{Kingma2017} with a learning rate of $3 \times 10^{-3}$. The model is optimized by minimizing the negative log-likelihood of the $M$–$R$ samples.

To evaluate, we compare the probability distribution learned by the normalizing flow to our test set using the Kullback-Leibler (KL) divergence, which provides a single number quantifying how closely the entire learned distribution approximates the target one. This results in KL divergences of 0.0075 and 0.0023 for the normalizing flow of the CS and PP EOS prior distribution, respectively. A KL divergence close to zero indicates that the distribution learned by the normalizing flow is nearly indistinguishable from the test set. See Fig.~\ref{fig:MR_priors} for a visual comparison between the prior model and the resultant normalizing flow. 

When sampling radii conditioned on a given mass, as in the cases of PSR J0740+6620 and PSR J0437-4715, samples with radii outside the interval 5.5 to 14.0 km for the PP EOS model and 6.5 to 14.5 km for the CS EOS model are excluded from the analysis. These bounds are chosen to avoid artifacts arising in regions where the normalizing flow assigns low but nonzero probability and to avoid drawing unphysical $M$–$R$ values. The bounds are conservative and lie well outside the support of the EOS prior. For a sample size of $1\times10^6$, the probability of drawing samples outside these bounds is close to zero, so truncation is not expected to significantly affect the normalization of the distribution. 

\section{Data}\label{sec:data}

In this section, we outline the x-ray event data used in this study. In Sec.~\ref{subsec:J0740data}, we describe the most important aspects of the dataset from the high-mass pulsar PSR J0740+6620, as detailed in \citet{Salmi2024-J0740}, and \citet{Hoogkamer2025-J0740}. Section~\ref{subsec:J0437data} gives a brief description of the observations from the nearest and brightest rotation-powered millisecond pulsar PSR J0437-4715, as detailed in \citet{Choudhury2024-J0437}. In Sec.~\ref{subsec:EOS priors dist}, we summarize the prior distributions of the PP and CS EOS models.  

\subsection{PSR J0740+6620} \label{subsec:J0740data}

For PSR J0740+6620, we use the NICER dataset between September 21, 2018, and April 21, 2022, identical to \citet{Salmi2024-J0740, Hoogkamer2025-J0740}. We use the pulse-invariant channel subset [30, 150), corresponding to the nominal photon energy range [0.3, 1.5] keV. The data are processed in the same way and include 2.73 Ms (roughly a month) of on-source exposure time after filtering. For background constraints, the same XMM-Newton phase-averaged spectral data and blank-sky observations are used with the three EPIC instruments (pn, MOS1, MOS2), including the energy channels [57, 299) for pn, and [20, 100) for both MOS1 and MOS2 (as reported in \cite{Riley2021-J0740, Wolff2021, Salmi2022-J0740, Salmi2024-J0740, Hoogkamer2025-J0740}). Furthermore, the hot emitting regions of the neutron star are modeled with two independent circular uniform temperature spots using the \verb|ST-U| (\textit{Single Temperature-Unshared}) model (see \cite{Riley2019-J0030-xpsi, Vinciguerra2023-J0030-sim} for more details on the X-PSI model naming convention and parameters).

\subsection{PSR J0437–4715} \label{subsec:J0437data}

Similar to the headline result of PSR J0437-4715 in \citet{Choudhury2024-J0437}, we use the 3C50 NICER dataset, which contains observations from July 6, 2017 to October 11, 2021. This dataset owes its name to the usage of the 3C50 background model, which is designed to estimate the instrumental background in NICER observations. The model utilizes NICER event data to predict the background level as a function of the energy (see \cite{Remillard2022-3C50} for the model definition). To get the best estimate of the instrumental background of NICER, the data are filtered according to specific criteria (see \cite{Salmi2022-J0740} for a detailed description of the analysis and filtering procedure). The filtering steps reduce the usable exposure time from 2.736 to 1.328 Ms, but enable the quantification of the uncertainties in the net background spectrum, which is predominantly influenced by systematic instrumental effects. 

To maximize the signal-to-noise ratio and mitigate contamination from the bright Seyfert II active galactic nucleus RX J0437-4711, located $4^{\prime}.18$ from PSR J0437-4715, NICER observations were conducted pointing $1^{\prime}.15$ southwest of the pulsar. This deliberate offset reduces the AGN flux within the field of view to only $\sim$2.25\% (i.e., $\sim0.2$ counts/s). The AGN flux is then modeled as an additive background component. 

For the analysis, only the events in the pulse-invariant channels [30, 300) corresponding to the nominal photon energy range 0.3–30 keV are kept. This is done to match the energy range from XMM-Newton observations, which are used to cross-validate the results from NICER. Lastly, the hot emitting regions of the neutron star are modeled with a two-temperature spot and a ring using the \verb|CST+PDT| (\textit{Concentric Single Temperature + Protruding Dual Temperature} model). The two-temperature hot spot consists of two overlapping spherical emitting regions, referred to as the \textit{ceding} and \textit{superseding} components. The ring is modeled with two spherical components with one emitting and one masking (the latter referred to as \textit{omitting} in \cite{Vinciguerra2023-J0030-sim, Vinciguerra2024-J0030}).   

\subsection{PP and CS EOS prior distributions}\label{subsec:EOS priors dist}
With the EOS-informed PPM framework in place (Sec.~\ref{subsec:PPM EOS prior and EOS inference}), the next step is to compute the $M$–$R$ priors for the PP and CS EOS models. This involves constructing the EOS models across the relevant density range and defining the prior ranges that specify their distributions. Following \citet{Rutherford24}, we briefly summarize the EOS construction adopted in this work and the corresponding priors.

To construct both the PP and CS EOS models, we adopt the BPS crust EOS \cite{Baym71} for densities $\leq 0.5 n_0$, with log-linear interpolation to the first $\chi$EFT data points at $0.5792 n_0$. For densities in the range $0.5792 n_0 \leq n \leq 1.5 n_0$, we rely on N$^3$LO $\chi$EFT calculations from \citet{Keller2023} of neutron star matter in beta equilibrium. Above $1.5 n_0$, we extend the EOS to high densities using the PP or CS parametrizations. 

The priors on the N$^3$LO $\chi$EFT band are determined by a single polytropic fit, i.e., $P(n) = K (n/n_0)^\Gamma$, to the upper and lower pressure bounds for $n \in [0.5, 1.5] n_0$. This fit yields prior ranges of $K \in [2.207,3.056] \, \mathrm{MeV} \, \mathrm{fm}^{-3}$ and $\Gamma \in [2.361,2.814]$ \cite{Rutherford24}. 

For the PP model, the priors are defined by varying three polytropic indices, $\Gamma_1$, $\Gamma_2$, and $\Gamma_3$, as well as two transition densities, $n_{12}$ and $n_{23}$, which connect the first to the second polytrope and the second to the third, respectively. The priors on these parameters are given by: $\Gamma_1 \in [0,8]$, $\Gamma_2 \in [0,8]$ $\Gamma_3 \in [0.5,8]$, and $2 n_0 \leq n_{11} \leq n_{23} \leq 8.3 n_0$. 

For the CS model, the prior distribution is defined by parametrizing the speed of sound $c_s^2 = dP/d\varepsilon$, using five parameters $a_1$ to $a_5$, which are varied over the following ranges: $a_1 \in [0.1,1.5]$, $a_2 \in [1.5,12]$, $a_3/a_2 \in [0.05,2]$, $a_4 \in [1.5,37]$, and $a_5 \in [0.1,1]$. The resulting prior distributions for both the PP and CS models span a large range of possible EOSs. For additional details on the PP and CS EOS models, see \cite{Rutherford24,Hebeler2013,Greif19}

Finally, all EOS priors and posteriors are required to obey causality and the constraint that neutron stars have masses $M\geq 1$ \Msun. This lower mass limit is imposed to be consistent with the lowest confirmed neutron star mass observation \cite{Martinez:2015mya} and is also motivated by theoretical models of early neutron star evolution \cite{Strobel_1999} as well as the minimum remnant masses predicted by simulations of core-collapse supernovae \cite{Janka2008,Fischer2010,Radice2017,Suwa2018,Muller2025}. For completeness, the resulting $M$–$R$ prior distributions of both of the PP and CS models are shown in Fig.~\ref{fig:MR_priors}. 

\section{Results} \label{sec:Results}

The results, aimed at showing the impact of incorporating the EOS prior information in the analysis, are presented in two parts. In Sec.~\ref{subsec:ppm results}, we show the PPM inference results using the EOS $M$–$R$ prior information for both PSR J0740+6620 and PSR J0437-4715. In Sec.~\ref{subsec:EOS inference results}, we present the results of the subsequent EOS inference using the posteriors from Sec.~\ref{subsec:ppm results}. 

\subsection{PPM including EOS prior information}\label{subsec:ppm results}

To assess the influence of an EOS-informed $M–R$ prior on the inferred properties of PSR J0740+6620 and PSR J0437-4715, we evaluate two representative EOS models as priors: CS and PP. These are compared against the baseline EOS-agnostic approach.

\subsubsection{PSR J0740+6620}

\begin{figure}[t!]
\centering
\includegraphics[width=\columnwidth]
{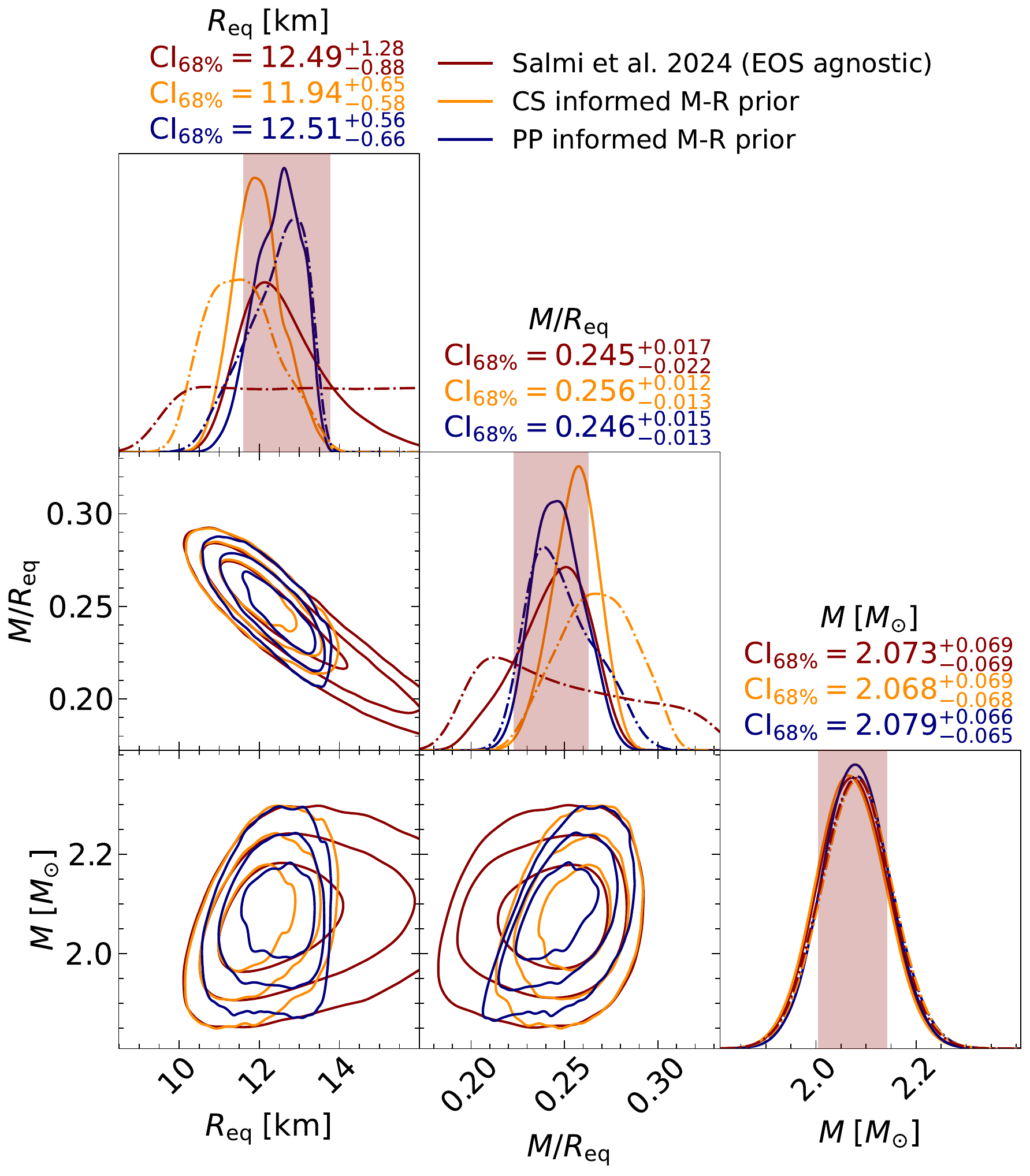}  
\caption{Radius, compactness, and mass posterior distributions using the PSR J0740+6620 joint NICER and XMM-Newton dataset conditional on the \texttt{ST-U} model. Three posterior distributions are shown: the results from \citet{Salmi2024-J0740} without an EOS-informed prior (red), the results obtained with the CS-informed $M$–$R$ prior (orange), and the ones obtained with the PP-informed $M$–$R$ prior (blue). All inference results are obtained using \texttt{MultiNest} with $4\times10^4$ live points and a sampling efficiency of 0.01. The marginal prior probability distribution functions for each parameter are displayed as dashed-dotted lines. The shaded regions in the diagonal panels contain the 68.3\% credible interval for each parameter symmetric around the median. The contours in the off-diagonal panels contain the 68.3\%, 95.4\%, and 99.7\% credible regions.}
\label{fig:corner_J0740}
\end{figure}

For PSR J0740+6620, we take the headline results of \citet{Salmi2024-J0740}, using the EOS-agnostic approach, as our baseline. The analysis of \citet{Salmi2024-J0740}, using \texttt{MultiNest} with $4\times10^4$ live points and a sampling efficiency of 0.01, reported an equatorial radius of $12.49^{+1.28}_{-0.88}$ km and a gravitational mass of $2.073^{+0.069}_{-0.069}$ \Msun. Figure~\ref{fig:corner_J0740} shows the posterior distributions, as well as the ones obtained when incorporating the CS and PP models as prior information on $M–R$. Table~\ref{tab:kl-divergence J0740} shows the accompanying KL divergence (in bits) between the posterior and corresponding prior for each model. 

The radius credible intervals for all three models overlap. However, the radius inferred using the CS-informed prior is $11.94^{+0.65}_{-0.58}$ km---with the median smaller than the baseline---while the PP-informed prior yielded a slightly larger median radius of $12.51^{+0.56}_{-0.66}$ km. The results are in line with our expectation, since the CS model favors softer EOSs and the PP model favors stiffer ones. However, once several sources are included, the CS and PP differences are smaller \cite{Rutherford24}. Notably, the mass and radius credible intervals obtained with the EOS-informed priors are narrower than those obtained using the EOS-agnostic $M$–$R$ prior. 

The posteriors of the geometry parameters, as shown in Fig.~\ref{fig:corner_J0740_all} of the Appendix, match even better for both the EOS-agnostic approach and the EOS-informed prior runs, with overlapping credible intervals for all parameters. Furthermore, the difference between the maximum likelihood and Bayesian evidence among all three models is negligible.\footnote{The difference in maximum log-likelihood is approximately $0.1$ across all three cases, while the evidence differs by approximately $+0.51$ for the CS-informed run and $+0.24$ for the PP-informed run relative to the EOS-agnostic approach.} 

In terms of computational cost, the EOS-agnostic inference required 84k core hours, compared to 21k and 22k core hours for the runs using the CS-informed and PP-informed priors, respectively.

\begin{table}[t!]
\resizebox{\columnwidth}{!}{%
\begin{tabular}{llll}
\hline
\textbf{Parameter} & \textbf{\begin{tabular}[c]{@{}l@{}}Salmi et al. 2024 \\ (EOS agnostic)\end{tabular}} & \textbf{\begin{tabular}[c]{@{}l@{}}CS-informed \\ $M$-$R$ prior\end{tabular}} & \textbf{\begin{tabular}[c]{@{}l@{}}PP-informed \\ $M$-$R$ prior\end{tabular}} \\ \hline
$R_{\text{eq}} [\text{km}]$ & 0.64 & 0.36 & 0.11 \\
$M/R_{\text{eq}}$ & 0.72 & 0.51 & 0.14 \\
$M [$\Msun$]$ & 0.01 & 0.03 & 0.01 \\ \hline
\end{tabular}%
}
\caption{KL divergence (in bits) between the posterior and corresponding prior for each parameter, computed for the three different inferences for PSR J0740+6620 shown in Fig.~\ref{fig:corner_J0740}.}
\label{tab:kl-divergence J0740}
\end{table}

\subsubsection{PSR J0437-4715}

\begin{figure}[t!]
\centering
\includegraphics[width=\columnwidth]
{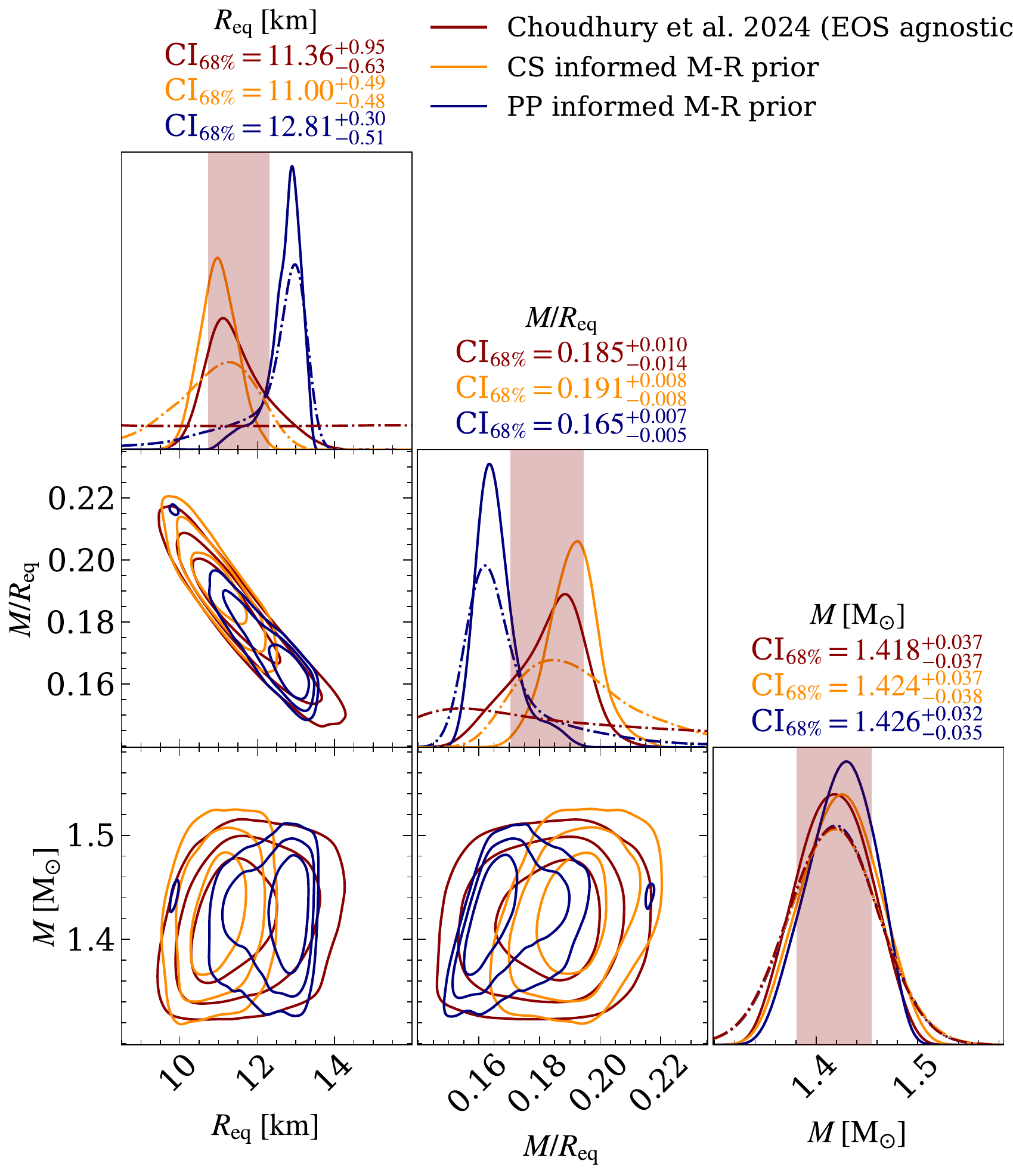}  
\caption{Radius, compactness, and mass posterior distributions using the PSR J0437-4715 NICER dataset conditional on the \texttt{CST+PDT} model. Three posterior distributions are shown: the headline result from \citet{Choudhury2024-J0437} using an EOS-agnostic approach (red), the results obtained with the CS-informed $M$–$R$ prior (orange), and the ones obtained with the PP-informed $M$–$R$ prior (blue). The EOS-agnostic analysis used \texttt{MultiNest} with $2\times10^4$ live points and a sampling efficiency of 0.3, while the EOS-informed runs used \texttt{MultiNest} with $4\times10^3$ live points and a sampling efficiency of 0.1. See Fig.~\ref{fig:corner_J0740} for more details about the figure elements.}
\label{fig:corner_J0437}
\end{figure}

For PSR J0437–4715, we adopted the headline result from \citet{Choudhury2024-J0437} as our baseline, using \texttt{MultiNest} with 20k live points and a sampling efficiency of 0.3. For the EOS-informed prior runs, we used $4\times10^3$ live points and a sampling efficiency of 0.1 to reduce computational costs. The resulting posterior distributions for all three models are shown in Fig.~\ref{fig:corner_J0437}. Table~\ref{tab:kl-divergence J0437} shows the accompanying KL divergence (in bits) between the posterior and corresponding prior. 

As shown in Fig.~\ref{fig:corner_J0437}, the EOS-agnostic approach yielded an equatorial radius of $11.36^{+0.95}_{-0.63}$ km and a gravitational mass of $1.418^{+0.037}_{-0.037}$ \Msun. The CS-informed prior run inferred a radius of $11.00^{+0.49}_{-0.48}$ km, which is slightly smaller than the baseline result, though the credible intervals otherwise strongly overlap. On the other hand, the PP-informed prior gives a radius of $12.81^{+0.30}_{-0.51}$ km, which is noticeably larger. Similar to PSR J0740+6620, these results align with expectations: The CS model favors softer EOSs given a smaller radius, while the PP model favors stiffer ones, giving a larger radius for the same mass. For both EOS modes, the mass and radius credible intervals are again narrower than those obtained from the EOS-agnostic approach.

\begin{figure*}[t!]
\centering
\includegraphics[width=0.8\textwidth]{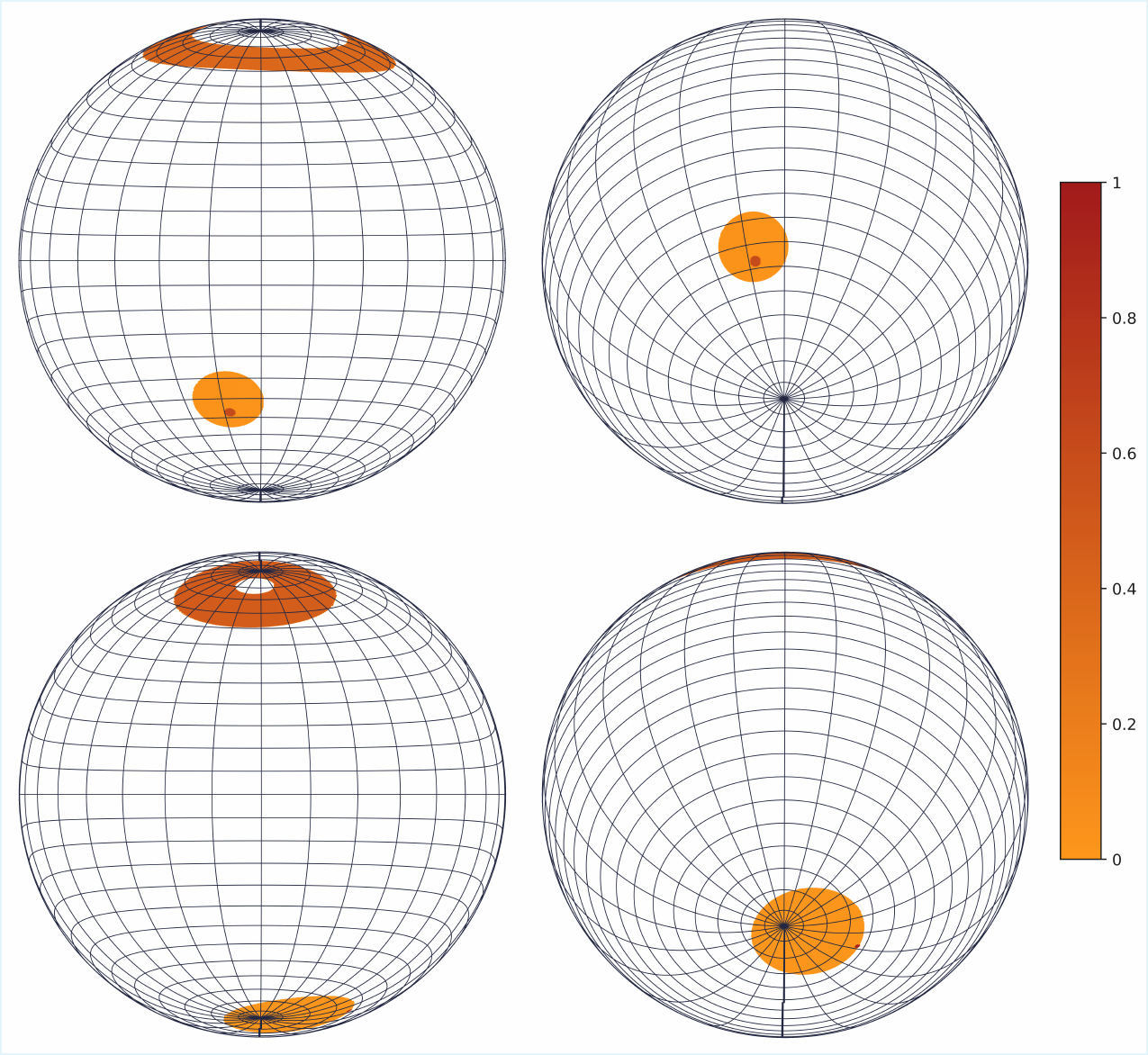}
\caption{Projections of the inferred maximum likelihood geometries are shown for the EOS-agnostic run of \citet{Choudhury2024-J0437} (top) and for the CS-informed $M–R$ prior run (bottom) using the PSR J0437–4715 NICER dataset, both conditional on the \texttt{CST+PDT} model. The projections are shown from the star's equatorial plane (left) and from the observer inclination (right). The color bar shows the normalized hot spot temperature, ranging between 0 (coolest, orange) and 1 (hottest, red). In both cases, the geometries consist of a ring surrounding the north pole and a two-temperature spot in the southern hemisphere. Note that in the CS-informed run (bottom), the small (superseding) hot spot in the southern hemisphere lies on the edge of the larger, cooler (ceding) region, whereas in the EOS-agnostic run (top) it is centered within the cooler region.}
\label{fig:projection plot J0437}
\end{figure*}

\begin{figure*}[t!]
\centering
\includegraphics[width=\textwidth]{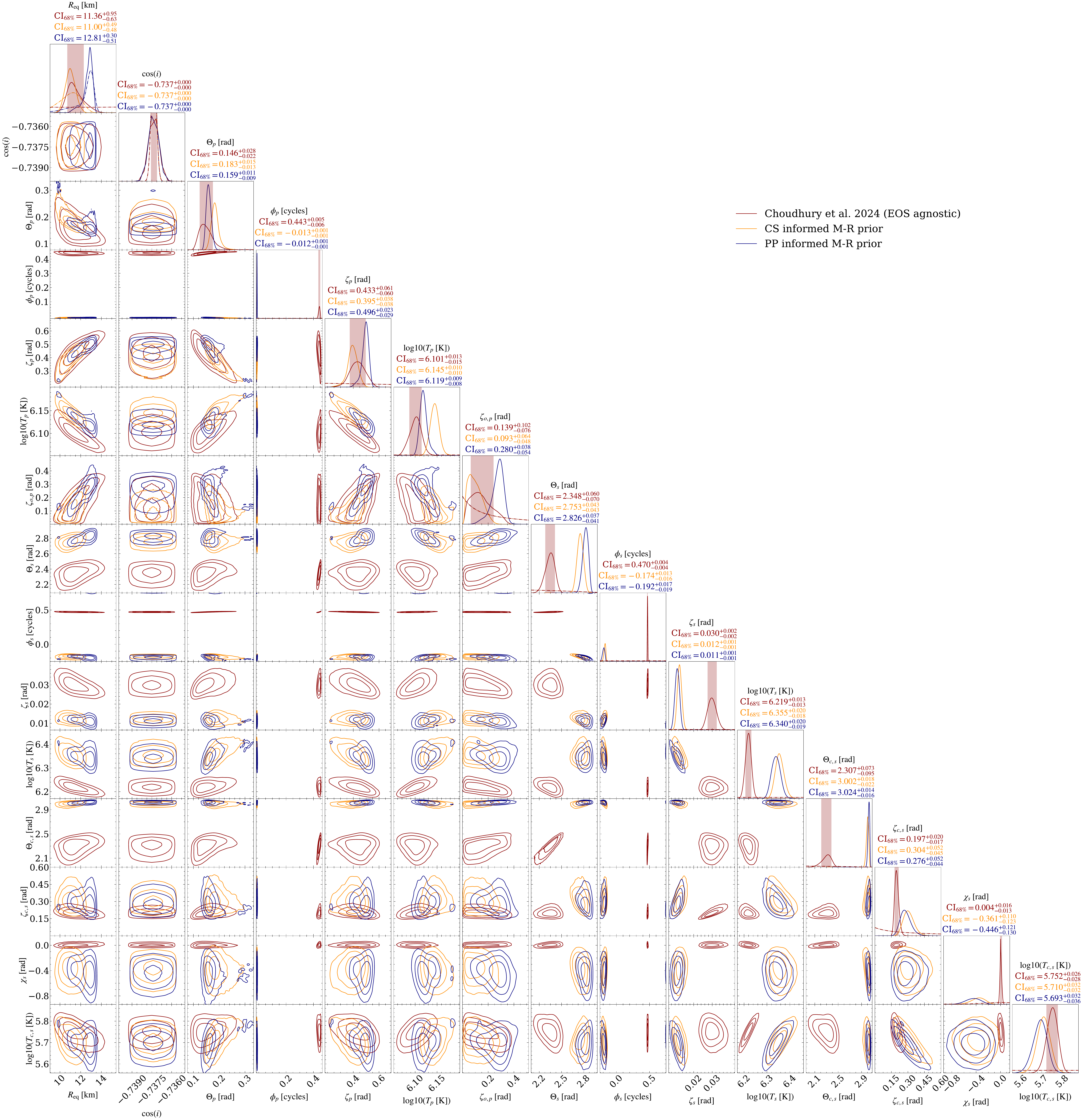}
\caption{Radius and spot geometry posterior distributions using the PSR J0437-4715 NICER dataset conditional on the \texttt{CST+PDT} model. Additional details of the parameter descriptions and the prior probability distribution functions can be found in Table A1 of \cite{Choudhury2024-J0437}.\footnote{The 68\% credible interval lower limit of the secondary superseding angular radius is -0.0023, not -0.023 as stated in that table.} See Fig.~\ref{fig:corner_J0437} for more details about the figure elements.}
\label{fig:corner_J0437_all}
\end{figure*}

For both the EOS-informed and EOS-agnostic cases, the inferred geometry consists of a ring surrounding the north pole---the primary hot spot with an omission region in the middle---and a two-temperature spot in the southern hemisphere, the secondary hot spot, consisting of a superseding and ceding region (see Fig.~\ref{fig:projection plot J0437} for the inferred maximum likelihood geometry). Notably, both the CS- and PP-informed prior runs yield geometric solutions that differ in hot spot size, temperature, and location from \citet{Choudhury2024-J0437}, as seen in Fig.~\ref{fig:corner_J0437_all}. Most credible intervals for the geometry parameters---especially those of the secondary hot spot---do not overlap, comparing the EOS-agnostic with the EOS-informed cases. The two EOS-informed runs are dominated by modes that are more extreme than the EOS-agnostic configuration, in terms of their surface emission patterns. In particular, the secondary superseding hot spot appears smaller, hotter, and closer to the south pole, while the ceding component is relatively larger, compared to both the superseding region and the EOS-agnostic solution. The temperature of the ceding component is similar for all cases. On average, the ring is comparable in size to the CS-informed run but becomes larger and thinner (with a larger omission region) in the PP-informed run, relative to the EOS-agnostic solution. For the maximum likelihood vector, the rings in both the CS- and PP-informed runs are smaller, slightly thicker with a smaller central omission region compared to the EOS-agnostic solution.

\begin{table}[t!]
\resizebox{\columnwidth}{!}{%
\begin{tabular}{llll}
\hline
\textbf{Parameter} & \textbf{\begin{tabular}[c]{@{}l@{}}Choudhury et al.\\ 2024 (EOS agnostic)\end{tabular}} & \textbf{\begin{tabular}[c]{@{}l@{}}CS-informed \\ $M$-$R$ prior\end{tabular}} & \textbf{\begin{tabular}[c]{@{}l@{}}PP-informed \\ $M$-$R$ prior\end{tabular}} \\ \hline
$R_{\text{eq}} [\text{km}]$ & 1.62 & 0.59 & 0.28 \\
$M/R_{\text{eq}}$ & 1.63 & 0.66 & 0.30 \\
$M [$\Msun$]$ & 0.08 & 0.07 & 0.16 \\ \hline
\end{tabular}%
}
\caption{KL divergence (in bits) between the posterior and corresponding prior for each parameter, computed for the three different inferences for PSR J0437-4715 shown in Fig.~\ref{fig:corner_J0437}.}
\label{tab:kl-divergence J0437}
\end{table}

Relative to the EOS-agnostic approach, the maximum likelihood and evidence from the CS-informed prior runs are both approximately $+38.5$ units higher, while those from the PP-informed prior runs are approximately $+38.6$ and $+34.2$ units higher, respectively, measured in natural logarithm. These differences in evidence correspond to a Bayes factor of $2\ln \text{BF}\approx 77.0$ (CS) and 68.4 (PP), constituting \textit{very strong} evidence in favor of the EOS-informed models, following the interpretation of \citet{Kass1995}. Furthermore, the CS-informed run is higher than the PP-informed run by $+4.3$ (evidence), yielding $2\ln \text{BF}\approx8.7$ which constitutes \textit{strong} evidence for the CS-informed model.

Regarding computational cost, the EOS-agnostic prior run required approximately 783k core hours, compared to 33k and 25k core hours for the runs using the CS-informed and PP-informed priors, respectively. Note that these runs are done with different sampler settings, which makes comparing the computational time difficult. 

\subsection{EOS inference}\label{subsec:EOS inference results}

We now examine the influence of using the EOS-informed PPM results, compared to the EOS-agnostic PPM results, for PSR J0740+6620 and PSR J0437-4715, as presented in Sec.~\ref{subsec:ppm results}, on the subsequent EOS inference for the PP and CS models. For the EOS inference only these PPM results are included and no additional astrophysics constraints, such as those from gravitational-wave observations, are used.

\begin{figure*}[t!]
\centering
\includegraphics[width=\textwidth]{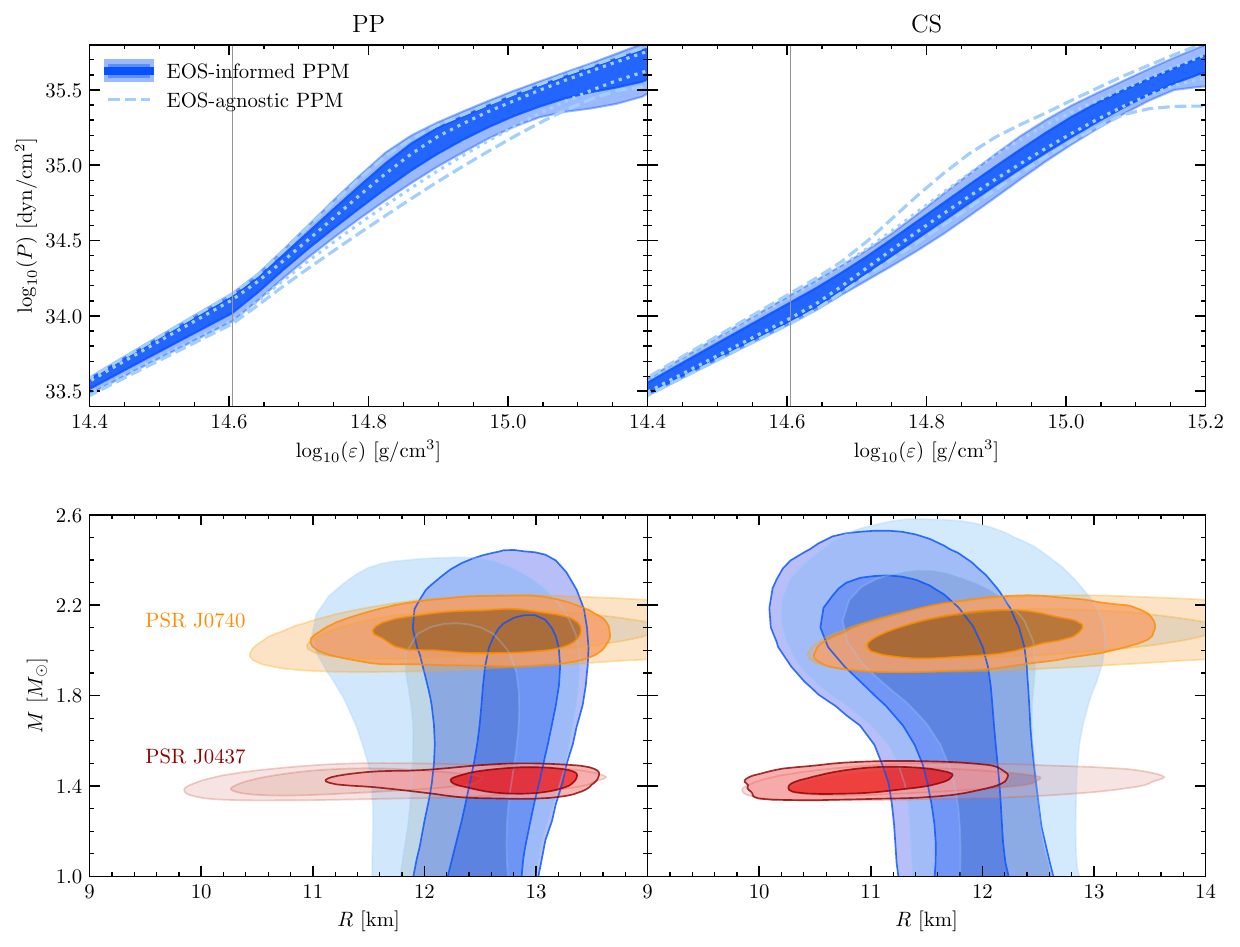}
\caption{Pressure-energy density (top) and mass-radius (bottom) posteriors for the PP (left) and the CS (right) model when using EOS-agnostic PPM results (dashed/dotted light blue lines/lightly shaded regions) and EOS-informed PPM results (dark blue shaded regions). In the mass-radius space, the EOS-informed PPM results for PSR J0740+6620 (orange shaded regions) and PSR J0437-4715 (red shaded regions) are dark shaded, while the EOS-agnostic PPM results are lightly shaded. The dark (light) and inner (outer) areas encompass the 68\% (95\%) credible regions. The vertical gray lines mark the transition density from the $\chi$EFT band to the high-density EOS extensions.}
\label{fig:pp and cs eos/mr posteriors}
\end{figure*}

Figure~\ref{fig:pp and cs eos/mr posteriors} shows the pressure-energy density and $M$–$R$ posteriors for both the PP and CS models obtained with the EOS-agnostic or the EOS-informed PPM results. For the PP model, the EOS-informed PPM results shift the pressure-energy density posterior toward stiffer EOSs, accommodating both the tighter radius constraint for PSR J0740+6620 and the larger radii inferred for PSR J0437-4715. Compared to the EOS-agnostic PPM results, the corresponding $M$–$R$ posteriors yield markedly narrower 68\% and 95\% credible regions, with an overall preference for larger stellar radii. On the other hand, the CS model with EOS-informed PPM results predicts relatively lower pressures at intermediate densities and correspondingly smaller radii, thereby favoring softer EOSs. Although the EOS-informed PPM results of the PP and CS models prefer stiffer and softer EOSs, respectively, both model posteriors exhibit some overlap. For instance, in the $M$-$R$ space, both EOS-informed PPM results overlap between 12 and 12.6 km. However, when all NICER sources and gravitational-wave constraints are included, the PP and CS posteriors overlap strongly, demonstrating the overall consistency between the two models by \citet{Rutherford24}.

\begin{figure}[t!]
\centering
\includegraphics[width=\columnwidth]{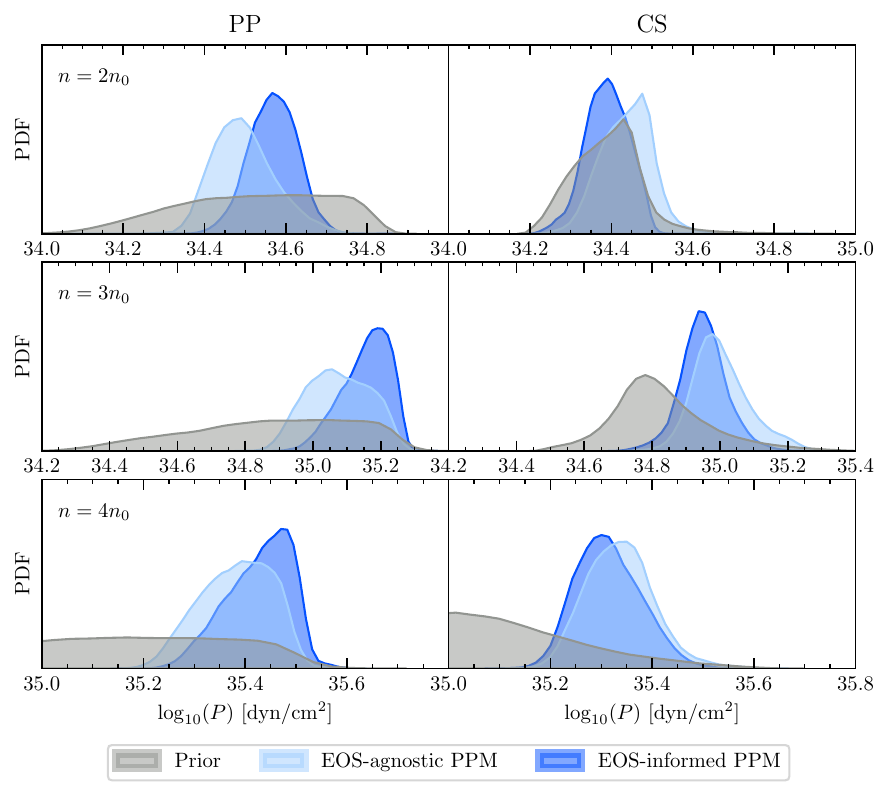}
\caption{Posterior (colored shaded regions) and prior (gray) pressure distributions at $2n_0$ (upper panels), $3n_0$ (middle panels), and $4n_0$ (lower panels). Results are shown for the PP (left) and CS (right) models using the EOS-agnostic (light blue) and EOS-informed (dark blue) PPM results.}
\label{fig:pp and cs pressure posteriors}
\end{figure}

Figure~\ref{fig:pp and cs pressure posteriors} shows the posterior pressure distributions at $2,3, \, \text{and} \,4n_0$, illustrating the impact of using EOS-informed PPM results compared to EOS-agnostic PPM results on intermediate-density EOS inferences. For the CS model, the pressures across all three fixed densities shift toward lower central values. On the other hand, for the PP model the pressure shifts toward higher values across all three densities, with narrower and more sharply peaked distributions at $3n_0$ and $4n_0$ when including EOS-informed PPM results compared to EOS-agnostic PPM results in the EOS inference. Similar to the results in Fig.~\ref{fig:pp and cs eos/mr posteriors}, the EOS-informed PPM pressure posteriors show some overlap across all three fixed densities between the PP and CS models.

\section{Discussion and future work} \label{sec:Discussion}

In this study, we implemented and tested the use of EOS-informed $M$–$R$ priors in the PPM pipeline and the effect of this on the subsequent EOS inference. We compared this approach to our original EOS-agnostic approach focusing on two pulsars, PSR J0740+6620 and PSR J0437-4715, and two EOS models based on $\chi$EFT calculations combined with a CS or PP extension to high densities.

\subsection{PPM including EOS prior information}

Our main goal was to implement the EOS-informed $M$–$R$ priors within the PPM framework and, in doing so, obtain tighter constraints on neutron star (geometry) parameters while reducing computational cost. This objective was successfully met: For both pulsars, the credible intervals derived with EOS-informed priors are narrower than those from the EOS-agnostic approach, and the posteriors behave as expected, with the CS (PP) model favoring somewhat smaller (larger) radii relative to the EOS-agnostic case. As anticipated, the computational costs for PSR J0740+6620 were significantly reduced when using EOS-informed priors, due to the smaller parameter space the sampler had to explore. For PSR J0437-4715, however, a direct comparison of computational costs could not be made because different sampler settings were used in the EOS-agnostic and EOS-informed runs in order to reduce run-time.

\subsubsection{PSR J0740+6620}

For PSR J0740+6620, the inferred masses for all three models are almost identical, with the mass posteriors being almost completely dominated by the mass prior from radio timing and the corresponding KL divergences close to zero. This outcome is as expected, given the mass prior from radio timing is highly informative. 

Regarding the radius, the credible intervals overlap for all three models. Nevertheless, the median radius obtained with the PP-informed $M$–$R$ prior resembles that from the EOS-agnostic run of \citet{Salmi2024-J0740} slightly better than the median radius from the CS-informed run. This is expected, since the PP radius prior overlaps more strongly with the EOS-agnostic posterior than the CS prior does, as seen in Fig. \ref{fig:corner_J0740}. 

In the case of the PP model, both the radius and the compactness are strongly prior dominated, with KL divergences of 0.11 and 0.14, respectively. Similar behavior is observed, though to a lesser extent, for the CS model, with a KL divergence of 0.36 for the radius and 0.51 for the compactness. Furthermore, for both EOS-informed runs, the posterior distributions extend to the upper bound of the radius prior and the lower bound of the compactness prior. Ideally, the physics-informed EOS model (e.g., by the $\chi$EFT constraint) is adapted so that the posterior is shaped primarily by the likelihood, ensuring that the inferred parameters are constrained by the data rather than the prior.

Apart from these differences, no particular model is clearly favored: The maximum likelihood and Bayesian evidence values for all three models are nearly identical across runs.

\subsubsection{PSR J0437-4715}
\label{J0437discussion}

For PSR J0437-4715, the inferred masses for all three models are again nearly identical, with the mass posteriors largely prior dominated and corresponding KL divergences around 0.1. Similar to PSR J0740+6620, this is expected given the highly informative mass prior from radio timing. 

Concerning the radius, the credible interval obtained with the CS-informed prior overlaps with the EOS-agnostic result from \citet{Choudhury2024-J0437} at the 68\% level, while the PP-informed prior only overlaps at the 95\% level, as shown in Fig.~\ref{fig:corner_J0437}. In both EOS-informed prior runs, the radius posterior is concentrated around the peak of the corresponding prior. However, for both the radius and the compactness posteriors, the data are clearly informative with KL divergences between 0.28 and 0.66, although the prior still contributes significantly. The EOS-agnostic model yields KL divergences of about 1.6 for both radius and compactness, indicating a substantial information gain from the data.

It is notable, however, that the CS- and PP-informed $M–R$ prior runs yield significantly different geometric solutions from the one obtained in the EOS-agnostic run of \citet{Choudhury2024-J0437}. In particular, the secondary dual temperature hot region is closer to the south pole in the EOS-informed runs, with the superseding hot spot smaller and hotter than in the EOS-agnostic case. This difference arises because the additional $M–R$ constraints imposed by the EOS models during the PPM stage guide the sampler to explore this region of the parameter space more thoroughly. These solutions are sufficiently well separated in parameter space from those of the EOS-agnostic runs that we can consider them to be a separate mode. 

\subsubsection{Statistical properties of PSR J0437-4715's geometry}

Before discussing the physical properties of the newly identified geometric mode, we first discuss its statistical properties. In this regard, the maximum likelihood and evidence values from the CS- and PP-informed prior runs are significantly higher than those obtained in the EOS-agnostic analysis from \citet{Choudhury2024-J0437}, with the CS-informed run yielding slightly larger values than the PP-informed run. To determine which mode is formally preferred in a statistical sense, however, requires both modes to be fully explored within the same converged run using the same prior.

To address this, we first revisited the EOS-agnostic run to check whether the new mode had been identified and explored in that analysis. Inspection of the samples from the EOS-agnostic run (from the Zenodo repository \cite{choudhury2024-Zenodo}), which did not use the multimode exploration setting, shows that \texttt{MultiNest} initially drew samples from regions corresponding to the CS- and PP-informed geometry modes. The sampler even continued to explore this part of parameter space to roughly 150 units below its maximum log-likelihood. However, as the run progressed, the sampler eventually concentrated on the geometric mode reported as the final result in \citet{Choudhury2024-J0437}.  

In principle, running with the multimode setting on should allow more thorough exploration of multiple different modes by locking in a select number of live points onto identified modes in proportion to the prior mass. In practice, however, this would increase the computational cost further still (the 20k live point run already required 783k core hours, and enabling multimode exploration requires even more live points to properly explore each identified mode). Carrying out such a run is therefore not possible with the available computational resources.  

Given these limitations, we instead attempted runs with the more restricted (and hence more computationally tractable) CS- and PP-informed priors using \texttt{MultiNest} with the multimode exploration setting turned on. However, for the CS-informed prior using 4k live points and a sampling efficiency of 0.3, this resulted in seven distinct modes. One mode resembled the EOS-agnostic solution, while the six remaining modes were similar to each other in terms of most, but not all, parameters of the CS-informed run with the multimode exploration turned off. 

By contrast, the PP-informed prior proved more tractable. With 8k live points and a sampling efficiency of 0.3, the multimode run identified two modes: one mode similar to the EOS-agnostic result, and one similar to the EOS-informed prior run, consistent within 95\% credible level for all parameters, and within 68\% credible level for most parameters. The maximum likelihood and local evidence values for the latter mode, corresponding to the more extreme geometric solution from the EOS-informed prior run, are higher by approximately +35.3 and +29.6 units, respectively, compared to the mode resembling the EOS-agnostic result. This difference in evidence corresponds to a Bayes factor of $2\ln \text{BF}\approx 59.1$, constituting \textit{very strong} evidence in favor of the new more extreme geometric mode found by the EOS-informed models, following the interpretation of \citet{Kass1995}. 

\subsubsection{Physical plausibility of PSR J0437-4715's geometry}

We now turn to the physical properties of the newly identified geometric mode and assess whether it is physically plausible and consistent with multiwavelength constraints from this source, while noting that more definitive conclusions on its feasibility will require testing against more detailed models and incorporating multiwavelength observational constraints \citep[see, e.g., recent modeling by][]{Petri25}. 

Both the geometric mode obtained in the EOS-agnostic run and the more extreme mode identified in the EOS-informed prior runs feature hot spot geometries composed of combinations of ringlike and circular structures. These configurations can be consistent with the so-called ``quadrudipolar'' magnetic field configuration \citep{Gralla17,Chen20}. For instance, the hot spot observed near the northern magnetic pole could plausibly be explained by a nearly aligned dipolar rotator,\footnote{Nearly aligned dipolar rotator refers to a neutron star whose magnetic and rotation axes are almost parallel, and whose magnetic field has a simple dipole configuration, implying two antipodal magnetic poles (polar caps).} provided that a fit accommodating an additional smaller spot within the larger ring is viable. In such nearly aligned configurations, the expected current heating pattern includes a compact spot near the magnetic pole---where super-Goldreich-Julian currents are sustained by pair production---and an encircling ring formed by the volume return current \citep{Philippov15,Gralla16}. 

In contrast, the very small hot (secondary) superseding spot observed in the newly identified mode at the southern pole is hard to reconcile with pulsar electrodynamics. Such narrow, localized structures are problematic, as they support only limited electric potential and are difficult to sustain through pair production processes \citep{Chernoglazov24}. In this respect, the somewhat larger superseding spot for the geometric mode identified in \citet{Choudhury2024-J0437} is more physically reasonable. 

To further assess whether the inferred hot spot geometry aligns with independent observational constraints, we next consider the viewing geometry of the neutron star as implied by radio observations. The geometrical priors used so far are based on the orbital inclination angle. For a narrow radio emission beam and a fully recycled millisecond pulsar binary, one would expect the orbital inclination angle to be close to the viewing angle between the magnetic axis and the observer's line of sight \citep[e.g.,][]{2021MNRAS.504.2094K}. However, the radio pulse of PSR J0437-4715 is very wide, with multiple connected components that are easily recognizable, spanning at least 200\degree\ in pulse longitude. With sufficient sensitivity, low-level radio emission can be detected covering roughly 330\degree\ of longitude. This can be explained either by a very wide radio beam, by a nearly aligned rotator (as in the new geometric mode)---which would appear to be inconsistent with the orbital inclination angle information---or by emission originating from a different source. 

In addition, we consider the gamma-ray emission as additional multiwavelength constraint on the inferred geometry. Recently, \citet{2025arXiv251005778K} argued that some radio emission of millisecond pulsars originates not from the polar cap but from beyond the light cylinder, congruent with gamma-ray emission. PSR J0437-4715 indeed shows prominent gamma-ray emission, with significant overlap in pulse phase between the radio and gamma-ray pulse profiles. As argued by \citet{2025arXiv251005778K}, this could explain both the wide longitude range of detected radio emission and the complex position angle swing that deviates from a rotating vector model. Solely considering the prominent gamma-ray emission detection suggests a combination of a large viewing angle and magnetic inclination angle, which would be inconsistent with an aligned rotator \citep[see, e.g.,][]{2025arXiv251005778K}. Fitting a rotating vector model is not only extremely difficult, but also inadequate in principle if part of the radio emission does not originate from the polar cap. 

A consistent picture is that both the viewing and magnetic inclination angles are large, the radio beam is intrinsically quite wide, and parts of the emission are likely originating from beyond the light cylinder. The geometrical information supplied by the orbital inclination angle is consistent with this image and remains consistent with the adopted priors. All in all, since the newly identified geometric mode in the EOS-informed prior run is close to a nearly aligned configuration, the less-aligned geometry obtained by the EOS-agnostic run is more plausible according to the gamma-ray and radio observations.

The different, more extreme geometry, also has an effect on the inferred background. The overall median background for the EOS-informed solutions is slightly lower than in the EOS-agnostic case (see the Zenodo repository$^{\ref{zenodo link}}$ for the background plots). This is consistent with the EOS-informed inferred geometry: The secondary ceding spot is relatively larger, and both the primary and secondary superseding components are (slightly) hotter, contributing more to the unpulsed emission of the pulse profile, which is otherwise compensated by the background in the EOS-agnostic case. Moreover, the secondary hot region (including both ceding and superseding components) lies closer to the south pole. Given the pulsar’s viewing angle, this spot remains more visible as the star rotates, further contributing to the unpulsed emission that would otherwise be compensated by the background. Lower and much better constrained backgrounds, as expected from upcoming telescopes eXTP \citep{LiA25} and NewAthena \citep{Cruise25}, should therefore help in identifying the correct geometric mode.

In conclusion, using EOS-informed priors in the PPM stage enabled the identification of a new mode, whose presence was already suggested by the EOS-agnostic run with multimode exploration setting off, but not fully explored. This mode, which has a more extreme surface emission pattern, is statistically favored over the EOS-agnostic solution by \citet{Choudhury2024-J0437}. Among the EOS-informed prior runs, the CS-informed run is preferred over the PP-informed run according to the Bayes factor, suggesting that the data likely favor smaller radii even for the new geometric mode, consistent with the EOS-agnostic result. Nonetheless, serious questions remain about whether the new, more extreme geometric configuration is physical, in particular with regard to the very small (secondary superseding) hot spot. Future work should address this issue and consider modifying the geometry priors to better reflect what is physically reasonable.

\subsubsection{Normalizing flow as EOS prior}

To model the EOS $M$–$R$ prior for the PPM step, we used a normalizing flow. We observed that the normalizing flow struggles to capture the $M$–$R$ distribution near $\approx 2.6$ \Msun ~and 13 km under the PP model, as shown in Fig.~\ref{fig:MR_priors}. This is likely due to sparse training samples at that boundary, which makes it more difficult for the neural network to learn the distribution edges. However, this is not a concern for our current analysis, since both pulsars have tightly constrained mass priors from radio timing which lie outside this region. We therefore do not sample from this part of the normalizing flow during the PPM step. 

\subsection{EOS inference}

The use of EOS priors in the PPM stage results in more stringent constraints on the inferred $M–R$ posteriors of PSR J0740+6620 and PSR J0437-4715. Including these posteriors in the EOS inference further tightens the PP (CS) EOS posteriors. In the $M–R$ plane, it produces a shift toward larger (smaller) radii and corresponding stiffening (softening) of the pressure-energy density relation for PP (CS), reflecting the expected behavior of these extensions, which relatively favor somewhat stiffer (softer) EOSs during the PPM inference. In particular, PSR J0437-4715 has the strongest impact on this effect. 

Studying the impact of EOS-informed PPM results on intermediate-density EOS inferences shows that, for the CS model, the pressures across all three densities consistently shift toward lower values. In contrast, for the PP model the pressures shift to higher values across all three densities. Moreover, at $3n_0$ and $4n_0$, the PP-informed pressure posteriors are more peaked than in the EOS-agnostic case. Overall, pressures increase for the PP model and decrease for the CS model at all densities, consistent with stiffening and softening of the EOS, respectively, when EOS-informed PPM data are used. It is interesting to note that in the EOS analysis with all NICER sources and including gravitational-wave constraints, the prior differences are significantly smaller \cite{Rutherford24}.

\subsection{Future work}

In future work, we aim to extend the analysis to sources such as PSR J1231-1411 \cite{Salmi2024-J1231} and PSR J0030+0451 \cite{Salmi2023-atmosphere, Vinciguerra2024-J0030} which exhibit complex, multimodal solutions that are challenging to sample thoroughly within reasonable computational time. A key question is whether incorporating EOS-informed $M$–$R$ priors during the PPM stage can help mitigate these problems.  

Moreover, we note that the current modeling framework is highly dependent on the choice of EOS parametrization. For more general applicability and reduced model bias, it would be advantageous to employ a more agnostic high-density EOS model, for example, based on Gaussian processes~\cite{Landry:2020vaw,Ng:2025wdj}. While empirical testing would be required to definitively address this, we anticipate that a sufficiently general high-density extension, such as that of a Gaussian process model, would likely converge toward the original EOS-agnostic framework. Nevertheless, the Gaussian process model may retain computational advantages through more efficient prior space constraints compared to uniform sampling over the full causal $M$–$R$ parameter space. The extent to which this produces tighter posteriors remains uncertain, particularly given that the broad PP and CS priors near 2.0\Msun~and~1.4\Msun~still yielded substantially more constrained posteriors compared to the EOS-agnostic approach for both PSR J0740+6620 and PSR J0437-4715, respectively.

This research has made use of data products and software provided by the High Energy Astrophysics Science Archive Research Center (HEASARC), which is a service of the Astrophysics Science Division at NASA/GSFC and the High Energy Astrophysics Division of the Smithsonian Astrophysical Observatory.

\textit{Software:} Cython \citep{Behnel2011}, Daft \citep{daft}, GetDist \citep{Lewis2019}, GNU Scientific Library \citep{Galassi2009}, HEASoft \citep{heasoft2014}, Matplotlib \citep{Hunter2007}, MPI for Python \citep{Dalcin2008}, \texttt{MultiNest} \citep{Feroz2009-Multinest}, nestcheck \citep{Higson2018-nestcheck-joss}, NumPy \citep{Walt2011}, \texttt{PyMultiNest} \citep{Buchner2016-PyMultiNest}, Pyro \cite{Pyro2018}, Python/C language \citep{Oliphant2007}, SciPy \citep{Jones2001-scipy}, NEoST \citep{Raaijmakers2025}, W\&B \citep{wandb}, and X-PSI \citep{xpsi}.

\textit{Facilities:} NICER, XMM-Newton

\begin{acknowledgments}
We are grateful to Sasha Philippov for providing input to Sec.~\ref{J0437discussion} on the physical implications of the newly identified geometric mode of PSR J0437-4715. M.H. and A.L.W. acknowledge support from the NWO grant ENW-XL OCENW.XL21.XL21.038 \textit{Probing the phase diagram of Quantum Chromodynamics} (PI: Watts). N.R. acknowledges support by NASA grant No.80NSSC22K0092. The work of N.R. was also supported in part by grant NSF PHY-2309135 to the Kavli Institute for Theoretical Physics (KITP). A.L.W. and D.C. acknowledge support from ERC Consolidator grant No.~865768 AEONS (PI: Watts). 
The work of M.M., I.S., A.S., and K.H. was supported by the European Research Council (ERC) under the European Union's Horizon 2020 research and innovation program (Grant Agreement No. 101020842).
T.S. acknowledges funding by the Research Council of Finland grant No.~368807.
The NWO Domain Science subsidized the use of the national computer facilities in this research.
Part of the work was also carried out on the HELIOS cluster including dedicated nodes funded via the above-mentioned ERC Consolidator grant.
M.H. and N.R. contributed equally to this work and are the joint lead authors. 

\end{acknowledgments}

\bibliography{main}

\appendix*
\section{Supplementary posterior plot for PSR J0740+6620} \label{sec:appendix J0740 corner}

\begin{figure*}
\centering
\includegraphics[width=\textwidth]{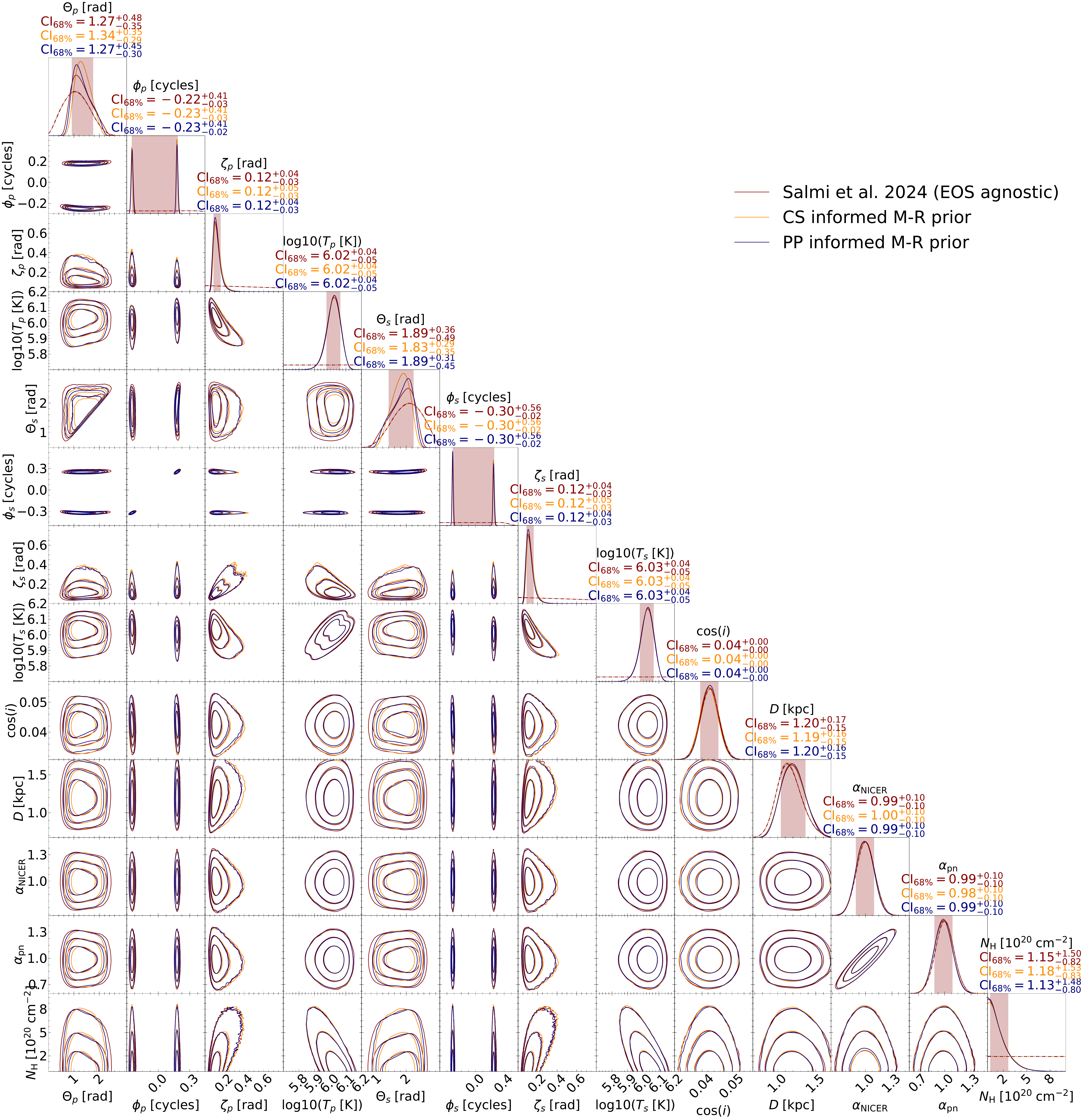}
\caption{Spot geometry and other parameter posterior distributions using the PSR J0740+6620 NICER and XMM-Newton dataset conditional on the \texttt{ST-U} model. See Fig.~\ref{fig:corner_J0740} for more details about the figure elements.}
\label{fig:corner_J0740_all}
\end{figure*}

Figure~\ref{fig:corner_J0740_all} shows the spot geometry and other parameter posterior distributions for PSR J0740+6620.

\end{document}